\documentclass[aps,twocolumn,secnumarabic,nobalancelastpage,longbibliography,amsmath,amssymb,nofootinbib,superscriptaddress]{revtex4-2}

\usepackage{graphics}      
\usepackage{graphicx}      
\usepackage{longtable}     
\usepackage{url}           
\usepackage{bm}            
\usepackage[dvipsnames]{xcolor}
\usepackage{verbatim}      
\usepackage{diagbox} 
\usepackage{xcolor}
\usepackage{hyperref}
\usepackage{tabularx}
\hypersetup{
    colorlinks=true,
    linkcolor=blue,
    filecolor=blue,      
    urlcolor=blue,
    citecolor=blue
    }
\newcolumntype{Y}{>{\arraybackslash}X} 

\begin{document}

\newcommand{\bea}{\begin{eqnarray}}
\newcommand{\eea}{\end{eqnarray}}
\newcommand{\be}{\begin{equation}}
\newcommand{\ee}{\end{equation}}
\newcommand{\bi}{\begin{itemize}}
\newcommand{\ei}{\end{itemize}}
\newcommand{\ds}{\displaystyle}
\newcommand{\rr}{{\bf r}}
\newcommand{\kk}{{\bf k}}
\newcommand{\pp}{{\bf p}}
\newcommand{\qq}{{\bf q}}
\newcommand{\uu}{{\bf u}}
\newcommand{\RR}{{\bf R}}
\newcommand{\vn}{{\bf 0}}
\newcommand{\ra}{\rangle}
\newcommand{\la}{\langle}
\newcommand{\sip}{{\sigma'}}
\newcommand{\ktyp}{k_{\rm typ}}
\newcommand{\pa}{\partial}
\newcommand{\vv}{{\bf v}}
\newcommand{\gr}{{\bf \nabla}}
\newcommand{\lb}{\lambda}
\newcommand{\up}{\uparrow}
\newcommand{\down}{\downarrow}
\newcommand{\Hr}{\mathcal{H}}
\newcommand{\Ar}{\mathcal{A}}
\newcommand{\Br}{\mathcal{B}}
\newcommand{\Vr}{\mathcal{V}}
\newcommand{\rP}{\mathcal{P}}
\newcommand{\Cr}{\mathcal{C}}
\newcommand{\Tr}{\mathcal{T}}
\newcommand{\Zr}{\mathcal{Z}}
\newcommand{\Dr}{\mathcal{D}}
\newcommand{\Sr}{\mathcal{S}}
\newcommand{\Nr}{\mathcal{N}}
\newcommand{\Mr}{\mathcal{M}}
\newcommand{\Ur}{\mathcal{U}}
\newcommand{\Pm}{\mathbb{P}}
\newcommand{\ddelta}{{\bf \delta}}
\newcommand{\Oom}{{\bf \Omega}}
\newcommand{\oo}{\Omega_1}
\newcommand{\ot}{\Omega_2}
\newcommand{\dt}{\frac{d}{dt}}
\newcommand{\ddt}[1]{\frac{d#1}{dt}}
\newcommand{\dz}{\frac{\partial}{\partial z}}
\newcommand{\ddz}[1]{\frac{\partial #1}{\partial z}}
\newcommand{\vel}{ v}
\newcommand{\etal}{{\it{et~al.}}}
\newcommand{\ket}[1]{\left| #1 \right>} 
\newcommand{\bra}[1]{\left< #1 \right|} 
\newcommand{\braket}[2]{\left< #1 \vphantom{#2} \right| \left. #2 \vphantom{#1} \right>} 
\newcommand{\matrixel}[3]{\left< #1 \vphantom{#2#3} \right| #2 \left| #3 \vphantom{#1#2} \right>} 
\newcommand{\for}{\mathcal F}
\newcommand{\dif}{\mathcal D}
\newcommand{\pdis}{\mathcal P}
\newcommand{\isat}{I_{\rm{sat}}}
\newcommand{\delt}{\delta^{\left|\mbox{\tiny $2'$}\right>}}
\newcommand{\delth}{\delta^{\left|\mbox{\tiny $3'$}\right>}}
\newcommand{\gamt}{\Gamma^{\left|\mbox{\tiny $2'$}\right>}}
\newcommand{\gamth}{\Gamma^{\left|\mbox{\tiny $3'$}\right>}}
\newcommand{\7}{$^7$Li}
\newcommand{\6}{$^6$Li}
\newcommand{\kb}{k_{\rm B}}
\newcommand{\ef}{E_{\rm F}}
\newcommand{\kf}{k_{\rm F}}
\newcommand{\tf}{T_{\rm F}}
\newcommand{\tc}{T_{\rm c}}
\newcommand{\tcb}{T_{\rm c,b}}
\newcommand{\tcf}{T_{\rm c,f}}
\newcommand{\om}{\omega}
\newcommand{\vext}{V_{\rm ext}}
\newcommand{\eb}{E_{\rm b}}
\newcommand{\bfit}[1]{{\fontfamily{ptm}\selectfont {\textit{\textbf{#1}}}}}
\newcommand{\pard}[2]{\frac{\partial#1}{\partial#2}}
\newcommand{\dd}[2]{\frac{d#1}{d#2}}
\newcommand{\bx}{{\bfit{x}}}
\newcommand{\ldb}{\lambda_{\rm dB}}
\newcommand{\avg}[1]{\langle#1\rangle}

\newcommand\blankpage{%
    \null
    \thispagestyle{empty}%
    \addtocounter{page}{-1}%
    \newpage}

\renewcommand{\thefootnote}{(\alph{footnote})}

\def\thefootnote{$^\ddagger$}\footnotetext{These authors contributed equally to this work.}

\title{A Multi-Purpose Platform for Analog Quantum Simulation}
\author{Shuwei Jin\thefootnote{}}
\altaffiliation{Present address: Physikalisches Institut, Universit\"at Heidelberg, Im Neuenheimer Feld 226, 69120, Heidelberg, Germany}
\affiliation{Laboratoire Kastler Brossel, ENS-Universit\'{e} PSL, CNRS, Sorbonne Universit\'{e}, Coll\`{e}ge de France, 24 rue Lhomond, 75005, Paris, France}
\author{Kunlun Dai\thefootnote{}}
\affiliation{Laboratoire Kastler Brossel, ENS-Universit\'{e} PSL, CNRS, Sorbonne Universit\'{e}, Coll\`{e}ge de France, 24 rue Lhomond, 75005, Paris, France}
\author{Joris Verstraten\thefootnote{}}
\affiliation{Laboratoire Kastler Brossel, ENS-Universit\'{e} PSL, CNRS, Sorbonne Universit\'{e}, Coll\`{e}ge de France, 24 rue Lhomond, 75005, Paris, France}
\author{Maxime Dixmerias}
\affiliation{Laboratoire Kastler Brossel, ENS-Universit\'{e} PSL, CNRS, Sorbonne Universit\'{e}, Coll\`{e}ge de France, 24 rue Lhomond, 75005, Paris, France}
\author{{Ragheed Alhyder}}
\altaffiliation{Present address: Institute of Science and Technology Austria (ISTA), Am Campus 1, 3400 Klosterneuburg, Austria}
\affiliation{Laboratoire Kastler Brossel, ENS-Universit\'{e} PSL, CNRS, Sorbonne Universit\'{e}, Coll\`{e}ge de France, 24 rue Lhomond, 75005, Paris, France}
\author{\\Christophe Salomon}
\affiliation{Laboratoire Kastler Brossel, ENS-Universit\'{e} PSL, CNRS, Sorbonne Universit\'{e}, Coll\`{e}ge de France, 24 rue Lhomond, 75005, Paris, France}

\author{Bruno Peaudecerf}
\affiliation{Laboratoire Kastler Brossel, ENS-Universit\'{e} PSL, CNRS, Sorbonne Universit\'{e}, Coll\`{e}ge de France, 24 rue Lhomond, 75005, Paris, France}
\affiliation{Laboratoire Collisions Agr\'egats R\'eactivit\'e, UMR 5589, FERMI, UT3, Universit\'e de Toulouse, CNRS, 118 Route de Narbonne, 31062, Toulouse CEDEX 09, France}

\author{Tim de Jongh}
\affiliation{Laboratoire Kastler Brossel, ENS-Universit\'{e} PSL, CNRS, Sorbonne Universit\'{e}, Coll\`{e}ge de France, 24 rue Lhomond, 75005, Paris, France}

\author{Tarik Yefsah}
\affiliation{Laboratoire Kastler Brossel, ENS-Universit\'{e} PSL, CNRS, Sorbonne Universit\'{e}, Coll\`{e}ge de France, 24 rue Lhomond, 75005, Paris, France}

\date{\today}

\begin{abstract}
Atom-based quantum simulators have had many successes in tackling challenging quantum many-body problems, owing to the precise and dynamical control that they provide over the systems' parameters. They are, however, often optimized to address a specific type of problems. Here, we present the design and implementation of a $^6$Li-based quantum gas platform that provides wide-ranging capabilities and is able to address a variety of quantum many-body problems. Our two-chamber architecture relies on a robust combination of gray molasses and optical transport from a laser-cooling chamber to a glass cell with excellent optical access. There, we first create unitary Fermi superfluids in a three-dimensional axially symmetric harmonic trap and characterize them using \textit{in situ} thermometry, reaching temperatures below 20\,nK. This allows us to enter the deep superfluid regime with samples of extreme diluteness, where the interparticle spacing is sufficiently large for direct single-atom imaging. Secondly, we generate optical lattice potentials with triangular and honeycomb geometry in which we study diffraction of molecular Bose-Einstein condensates, and show how going beyond the Kapitza-Dirac regime allows us to unambiguously distinguish between the two geometries. With the ability to probe quantum many-body physics in both discrete and continuous space, and its suitability for bulk and single-atom imaging, our setup represents an important step towards achieving a wide-scope quantum simulator.
\end{abstract}

\maketitle

\section{Introduction}

The last decades have seen the emergence of ultracold atom experiments as powerful platforms for quantum simulation of complex many-body systems \cite{Bloch2012,Daley2022review}. The success of atom-based quantum simulators stems from their ability to place a large number of particles in a well-characterized, tunable and isolated environment. For example, the energy landscape where particles evolve can be tailored to be uniform \cite{Navon2021}, harmonic \cite{Grimm2000}, periodic \cite{Gross2017}, disordered \cite{Choi2016,Garreau2017,Yan2017,Yan2017a,Abanin2019,MokhtariJazi2023}, or even tightly-confining in one or more directions to simulate a 1D or 2D system \cite{Bloch2008, Hadzibabic2009, DeDaniloff2021}. Interparticle interactions can be short range or long-range, repulsive or attractive, vanishingly weak or as strong as allowed by quantum mechanics \cite{Carr2009,Chin2010}. The atomic or molecular ensembles can be prepared in thermal equilibrium, out-of-equilibrium or be dynamically driven \cite{Goldman2014,Aidelsburger2022review}. The quantum-gas toolbox goes beyond these examples and offers many capabilities to create quantum systems with increasing complexity, which was underlined by major breakthroughs over the last two decades in our understanding of quantum matter \cite{Bloch2008,Bloch2012}, ranging from elucidating important properties of the BEC-BCS crossover \cite{Zwierlein2006,Nascimbene2010,Navon2010,Ku2012}, to the study of Bose and Fermi Hubbard Models \cite{Gross2017}, to the exploration of topological states of matter \cite{Zohar2015,Aidelsburger2022review}.

A limitation of current approaches, however, is that most quantum gas setups are optimized to address only a certain type of problems. One often makes a choice between lattice or continuous systems, short or long-range interactions, single-particle detection or bulk measurements, etc. In some cases, this is unavoidable as the experiment is designed to harness the properties of a specific atomic element or molecule. In many other cases, however, such constraints do not apply and combining experimental capabilities that had traditionally been used on distinct types of quantum many-body systems can have a useful impact on the ongoing quantum simulation effort.

Recently, promising steps were made in that direction: optical tweezers, for instance, have been combined with optical lattices in order to study single-atom quantum walks in two-dimensional lattices \cite{Young2022}, as well as with quantum gas microscopy to realize bottom-up quantum simulation of the Fermi-Hubbard model \cite{Spar2022}. Other recent work demonstrated the use of versatile energy landscapes on a given setup, for example, utilizing tunable tailored optical potentials to investigate dynamical symmetries of 2D Bose gases in continuous space under variable boundary conditions \cite{SaintJalm2019}, optical lattices with tunable geometry \cite{Tarruell2012,Kosch2022,Wei2023} or performing quantum gas microscopy of both triangular and square lattices within the same setup \cite{Yang2021}. 

\begin{figure*}
	\centering
	\includegraphics[width=\textwidth]{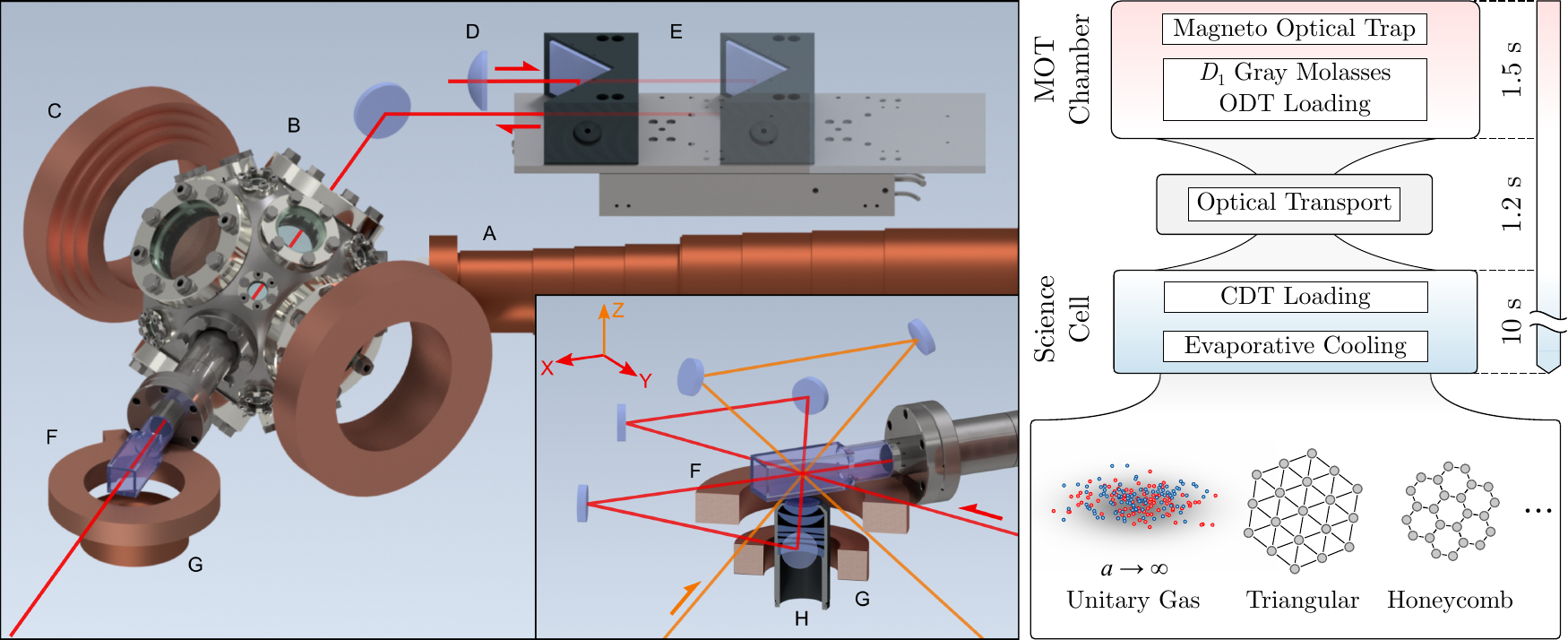}
	\caption{Overview of the experimental apparatus (left panel) and sequence (right panel). $^6$Li atoms slowed by a Zeeman slower (A) are collected in a magneto-optical trap (MOT) inside a spherical vacuum chamber (B), with a pair of coils (C) providing the magnetic gradient. Atoms are then cooled to sub-Doppler temperatures using a gray molasses, loaded into an optical dipole trap -- focused by a 1335\,mm lens (D) -- and subsequently transported into a glass cell using a linear translation stage (E). Two pairs of coils around the glass cell (F, G) are used to create the Feshbach bias field and apply a field curvature in the horizontal plane, respectively. The inset shows the glass cell with the horizontal lattice beam (red lines) and the vertical lattice beam (orange lines). Either a triangular or a honeycomb lattice geometry is generated by an appropriate choice of polarization, which is controlled by rotating a single half-waveplate (not shown) on the laser beam path at the entrance of the glass cell. A microscope objective (H) with numerical aperture of 0.55 is positioned below the glass cell for high-resolution imaging. Right panel: Laser cooling steps and their typical duration in the experimental sequence. Cooling to degeneracy typically takes 12 - 16\,s, after which we can study a variety of lattice and continuum systems.}
	\label{Fig:paper_figure}
\end{figure*}

Here, we present a multi-purpose quantum gas platform for the study of strongly-correlated Fermi systems in both lattice and continuous landscapes. Interacting fermionic systems play a special role among the various quantum many-body problems within reach of atom-based quantum simulators, as their understanding constitutes a serious challenge of modern physics. Indeed, theoretical approaches to tackling strongly-correlated fermionic systems are widely plagued by the infamous sign-problem, which limits the power of most unbiased numerical techniques in the thermodynamic limit, often forcing a resort to uncontrolled approximations, and in the past years the development of unbiased ‘sign-free’ approaches has been the subject of important ongoing efforts \cite{Muroya2003Review, Booth2009,CeperleyReview2010,Gandolfi2015ReviewNeutrons,Juillet2017,ZhangReview2017,Berger2021review,Alexandru2022RMP}. On the other hand, the experimental advances in quantum simulation not only have solved long-standing problems \cite{Zwerger2012,Zwierlein2013book,Yan2019, Mukherjee2019,Patel2020} but also helped the cross-validation of novel theoretical methods, such as the diagrammatic Monte Carlo methods \cite{Prokofev2007,Prokofev2008,Houcke2010,Houcke2012,Rossi2018, Houcke2019}. 

Our experiment is based on fermionic $^6$Li, which has proven to be a suitable atom to address a broad range of topics, from the BEC-BCS crossover \cite{Zwierlein2004,Zwierlein2005a,Zwierlein2006,Nascimbene2010, Ku2012, Valtolina2015, Mukherjee2017, Yan2019, Mukherjee2017,Patel2020, Kwon2020,Kwon2021}, to few-body and Efimov physics \cite{Ottenstein2008,Huckans2009,Williams2009,Nakajima2010,Lompe2010,Wenz2013,Zuern2013}, to quantum gas microscopy of lattice systems \cite{Omran2015,Parsons2016,Boll2016,Brown2019} including frustrated geometries \cite{Mongkolkiattichai2022}, to interacting Rydberg ensembles \cite{Guardado2021} and to novel cavity quantum electrodynamics effects, where photons couple to strongly-interacting matter \cite{Konishi2021}.

We use our setup to study three paradigmatic systems: the unitary Fermi gas in continuous space, the triangular lattice and the honeycomb lattice. Specifically, we first create unitary Fermi gases in a well-characterized trapping potential in three-dimensional continuous space and at controlled temperatures, which we obtain using \textit{in situ} thermometry based on state-of-the-art thermodynamics \cite{Ku2012,Zwierlein2013book}. Our coldest samples are deeply in the superfluid regime with absolute temperatures below 20\,nK and an average interparticle spacing of $\simeq1.3\,\mu$m, which brings their direct imaging via quantum gas microscopy within reach. Secondly, we present a versatile set of lattice configurations, which we characterize via matter-wave diffraction of a molecular Bose-Einstein condensate, in and beyond the Kapitza-Dirac regime. In particular, in the case of the triangular and honeycomb lattice geometries, we demonstrate how Bragg diffraction can be used to quantitatively discriminate between the two.

\section{Overview of the apparatus}

Our apparatus employs an all-optical strategy for producing deeply degenerate Fermi gases of $^6$Li atoms. A schematic overview is shown in Fig.\,\ref{Fig:paper_figure}. Its design is divided into two principal sections: (i) A preparation vacuum chamber in which the atoms are cooled to tens of microkelvins and loaded into an optical dipole trap (ODT), and (ii) a science glass cell to which the atoms are optically transported by moving the ODT, where they are evaporatively cooled to quantum degeneracy. This two-chamber structure provides good optical access to the science cell, which is crucial for our goal of tackling a diverse set of quantum problems within the same apparatus. In the following, we describe the general architecture of our setup and our experimental sequence. Additional experimental details can be found in Appendix~\ref{App:MOT}.

\subsection{Dipole trap loading}
\label{sec:dipole}

Our experimental sequence starts with a magneto-optical trap (MOT) of $^6$Li atoms loaded from a Zeeman-slowed atomic beam of lithium. After loading the MOT for typically 1.2\,s, we apply a compression stage (CMOT) by increasing the magnetic field gradient. We subsequently turn off the magnetic quadrupole field while keeping the MOT laser beams on for 3\,ms to hold the atoms until all transient magnetic fields have fully decayed and only residual static magnetic fields remain. We refer to this stage as the $D_2$ optical molasses phase. When all transient magnetic fields have decayed, we switch off the MOT beams, turn on the $D_1$ laser cooling beams for 3\,ms to capture the atoms in a gray molasses, and then reduce the intensity of the latter by half in 2 ms. A subsequent hold time of 1 ms allows the atoms to thermalize, and we obtain a cloud with phase space densities (PSD) of approximately $5\times10^{-5}$ and temperatures down to $40\,\mu$K which allows direct loading into an ODT (see Appendix~\ref{App:MOT} for details.). This strategy is similar to the one in Ref.\,\cite{Burchianti2014}.

\begin{figure}[t!]
	\centerline{\hspace{-0.\textwidth}
		\includegraphics[width=\columnwidth]{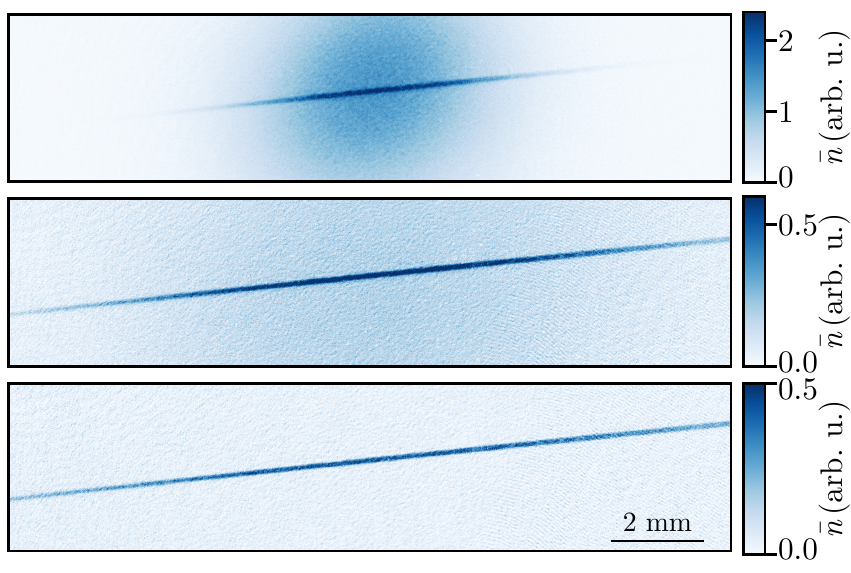}
	}
	\caption{Loading of ODT. Absorption images recorded on a CCD camera at various stages of the loading process for the single-beam ODT used to transport the atoms to the glass cell. Images taken after switching on the $D_1$ gray molasses for a duration of 5$\,$ms (top), 15$\,$ms (middle) and 30$\,$ms (bottom). The image axes are the physical camera axes, which are rotated with respect to the ODT propagation axis.}
	\label{Fig:IPGinMOTChamber}
\end{figure}

The (single-beam) ODT is turned on during the $D_2$ molasses stage and kept at maximal power of 156\,W throughout the gray molasses phase. It is derived from an Ytterbium fiber laser with a central wavelength of 1070 nm. At maximal power and with a beam waist of 90\,$\mu$m, we obtain a radial trapping frequency of $2\pi \times 4$\,kHz and a trap depth of $k_B\,\times\,$600\,$\mu$K, with $k_B$ the Boltzmann constant. With these parameters, we typically load $5\times10^6$ atoms in the $F = 1/2$ state in the ground state electronic manifold ($F$ being the hyperfine quantum number) at temperatures of 90$\,\mu$K, which remain nearly an order of magnitude smaller than the trap depth of the ODT. In Fig.\,\ref{Fig:IPGinMOTChamber} we show integrated density distributions $\bar{n}$ obtained by absorption imaging at different steps of this loading process.

\begin{figure}[t!]
	\centerline{
		\includegraphics[width=\columnwidth]{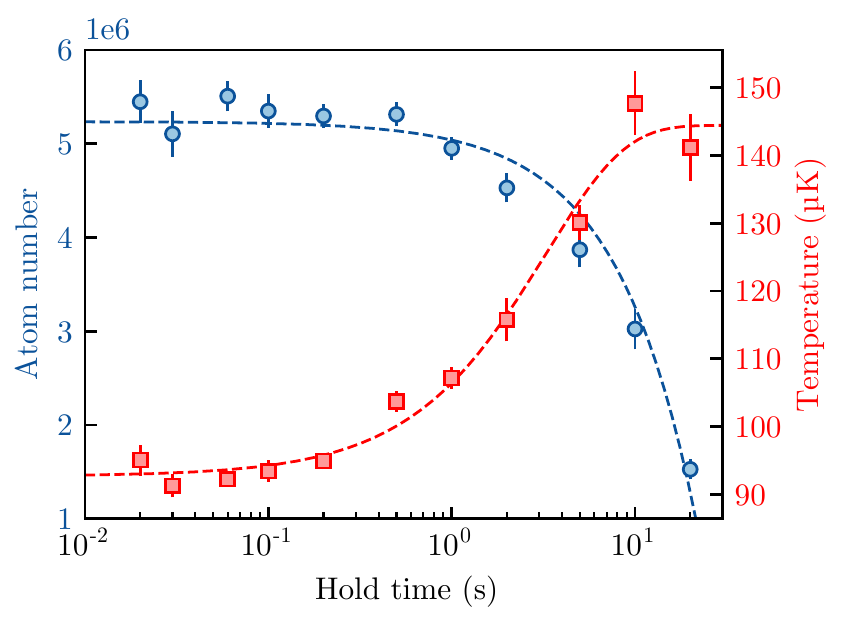}%
	}
	\caption{Number and radial temperature of atoms loaded in the transport beam dipole trap as a function of hold time in the ODT. Error bars denote one standard deviation of the mean from 15 measurements. The red dashed line is an exponential saturation curve which yields an initial heating rate of 16(3)\,$\mu$K/s. The blue dashed line is an exponential decay fit of the atom number.}
	\label{Fig:IPG_NT_vs_Holdtime}
\end{figure}

In anticipation of the optical transport to the science cell, we have studied the evolution of atom number and temperature while holding the atoms in the static ODT. The results are shown in Fig.\,\ref{Fig:IPG_NT_vs_Holdtime}, where radial temperature measurements are obtained from the expansion of the atomic distribution after a variable time-of-flight (TOF). We observe a slow heating of the atoms that results in their escape from the trap at long hold times. By fitting the evolution of the measured temperature with an exponential saturation curve, we extract an initial heating rate of 16(3)\,$\mu$K/s. This is larger than the estimated off-resonant photon scattering rate of 3.4$\,\mu$K/s and likely caused by low-frequency laser intensity noise. However, on the 1.2\,s timescale of transport to the glass cell, heating and atom loss are negligible.

\subsection{Transport and evaporative cooling}
\label{sec:evap}

We perform optical transport by shifting the ODT beam focus over a distance of 32\,cm in 1.2\,s. This is achieved using a motorized linear translation stage, carrying a pair of mirrors as shown in Fig. \ref{Fig:paper_figure}. In order to ensure a smooth motion and minimize heating and center-of-mass movement during transport, we define the trajectory to be a quartic function of time with a sigmoid shape, reaching a maximum velocity of 0.53\,m/s. The transport efficiency is larger than $97\%$ and yields samples of $5\,\times\,10^6$ atoms at a temperature of $125\,\mu$K in the glass cell (see Appendix \ref{App:MOT}).

At the glass cell the transport beam is crossed with a perpendicular laser beam at 1064\,nm, with a waist of $w = 61(1)\,\mu$m and a maximum power of 16\,W, forming a crossed dipole trap (CDT). The trapping potential due to the crossing beam initially has little influence with a trap depth of only $\sim100\,\mu$K. Then, the magnetic field is ramped in 200\,ms to $832\,$G corresponding to the center of  the broad Feshbach resonance between the two lowest hyperfine ground states of $^6$Li, which we denote $\ket{1}$ and $\ket{2}$. We prepare a spin-balanced sample of these two states by performing an adiabatic radio-frequency (RF) sweep spanning frequency values from far off-resonance to the resonant hyperfine transition frequency. We maintain the magnetic field at 832\,G, and hence strong resonant interactions, throughout the evaporation process to enhance elastic collision rates and achieve efficient evaporative cooling.

We initiate evaporation using a 2.8\,s exponential intensity ramp-down of the transport trap to 18\% of its maximum power, at which point its trap depth becomes comparable to the crossing dipole trap. Then we linearly decrease the intensity of both arms simultaneously down to a trap depth of approximately $k_B\,\times\,15\,\mu$K in 1.3\,s. The transport trap is subsequently switched off within 500\,ms with an exponential ramp-down. We end evaporation with a 0--4\,s linear intensity ramp-down of the laser beam perpendicular to the transport direction. This single ODT provides a strong (weak) confinement in the radial (axial) direction with a trap frequency of $2\pi \times 620\,{\rm Hz}\times\sqrt{P_\mathrm{ODT}/{\rm W}}$ ($2\pi \times 2.5\,{\rm Hz}\times\sqrt{P_\mathrm{ODT}/{\rm W}}$), with $P_\mathrm{ODT}$ the ODT power, while the magnetic coils provides additional axial trapping with a frequency of $2\pi\times 10$\,Hz, which largely dominates at low ODT power. \textit{In situ} absorption images taken during this evaporation step are shown in Fig.\,\ref{Fig:Evap}. At the lowest trap depth, we obtain samples of 5.1(1)$\times$10$^4$ atoms per spin state at temperatures of $17(1)\,$nK, well within the degenerate regime. The total cycle time for the production of a degenerate Fermi gas is 12 to 16\,s depending on the desired temperature (see Sec.\,\ref{sec:bulk}).

The optical access provided by the glass cell allows the degenerate sample to be loaded into a variety of different energy landscapes. We present below specific examples of how we can study continuous (Sec.\,\ref{sec:bulk}) and discrete (Sec.\,\ref{sec:lattice}) systems.

\section{Strongly-Interacting Fermions in Continuous Space}
\label{sec:bulk}

The two-component quantum-degenerate samples of fermions described above readily give access to the physics of strongly-interacting Fermi gases with tunable interparticle interactions and spin-population. Indeed, the broad Feshbach resonance between the two lowest hyperfine states of $^6$Li is ideally suited for the study of BEC--BCS crossover physics \cite{Zwerger2012} and more specifically the unitary regime, where the scattering length $a$ diverges. 

The unitary Fermi gas represents one of the critical challenges of quantum many-body physics \cite{Zwerger2012} and has been subject of major experimental and theoretical interest \cite{Bloch2012,Zwerger2012,Zwierlein2013book}, with relevance for the understanding of the dilute neutron matter in the crust of neutron stars, and the quark-gluon plasmas created in heavy ion collisions at $\sim10^{12}$\,K \cite{Schaefer2009, Patel2020}. Furthermore, because interactions do not introduce any energy or length scale (as a result of the diverging scattering length), it features remarkable universal properties \cite{Ho:2004a,Zwerger2012,Zwierlein2013book}. For instance, all its thermodynamic properties only depend on the ratio $T/T_F$, of the temperature $T$ to the Fermi temperature ${T_F = \frac{1}{k_B}\frac{\hbar^2}{2m}(3\pi^2n)^{2/3}}$ with $n$ the density of the cloud and $\hbar$ the Planck constant.

\begin{figure*}[t!]
	\centering
	\includegraphics[width=\textwidth]{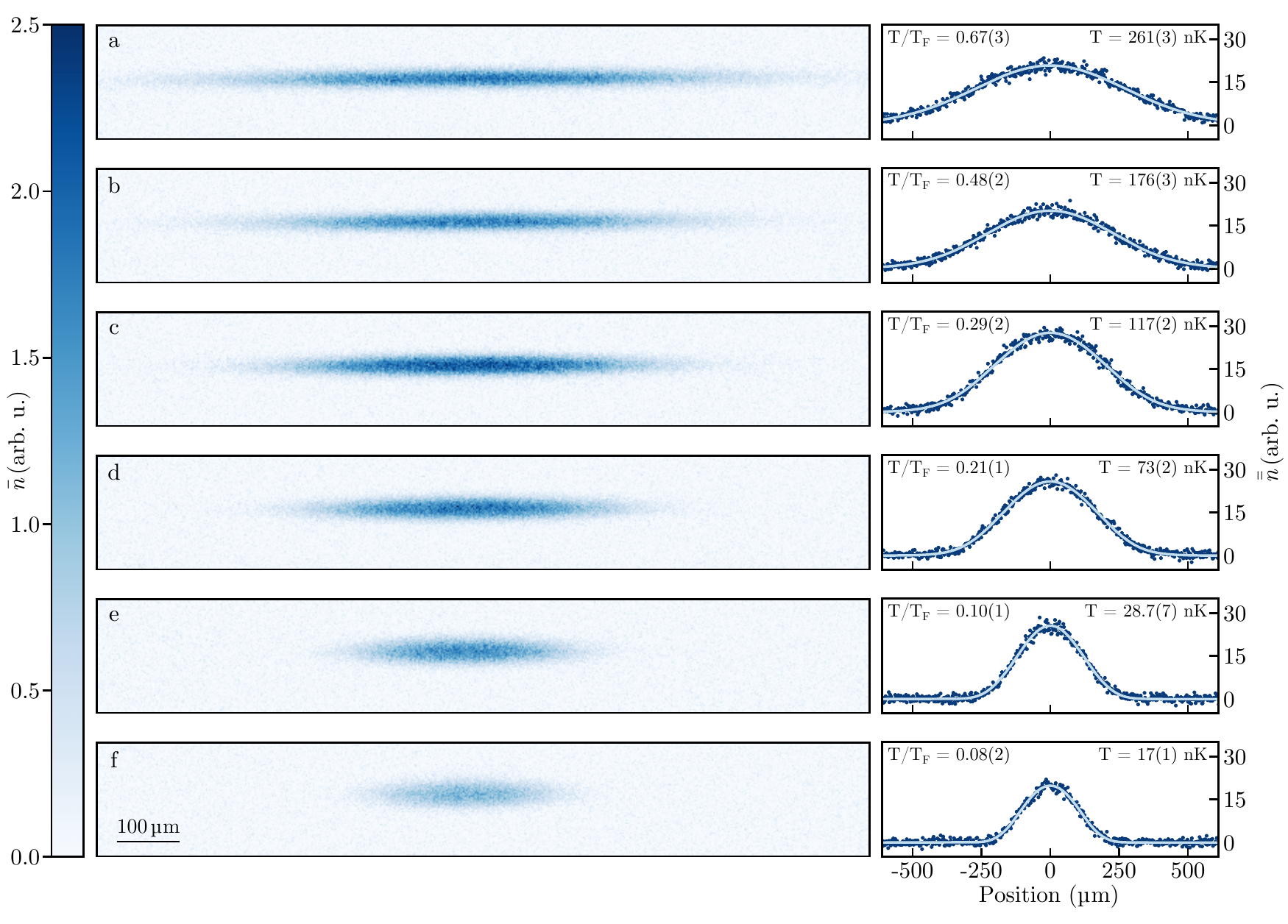}
	\caption{Thermometry of a unitary Fermi gas across the normal-to-superfluid transition. (Left) \textit{In situ} absorption images for decreasing final trap power: a) $300\,$mW; b) $200\,$mW; c) $125\,$mW; d) $75\,$mW; e) $30\,$mW; f) $20\,$mW. (Right) The corresponding doubly-integrated density profiles (blue data points) and fit to the EOS (light-blue curve), yielding absolute temperature $T$ and reduced temperature $T/T_{\rm F}$ at the center of the trap (both indicated in the respective panels).}
	\label{Fig:Evap}
\end{figure*}

Due to the strong interactions, standard thermometry relying on time-of-flight expansion does not apply to the unitary Fermi gas. Indeed, its expansion dynamics deviates from the ballistic behavior already at temperatures well above the superfluid transition temperature, and for many years, this hindered quantitative thermometry in the low temperature regime \cite{Ketterle2007, Zwierlein2013book}. In most experiments, degeneracy of strongly interacting fermionic gases is demonstrated using indirect or non-quantitative methods \cite{Ketterle2007, Zwierlein2013book}, such as the appearance of a bimodal density distribution in time-of-flight measurements performed on the BEC side of the Feshbach resonance.

Here, in order to reliably extract the temperature of our produced sample in the degenerate regime, we use the universal equation of state (EoS) of the homogeneous unitary Fermi gas for the pressure:
\begin{equation}
	P(\mu,T) = f(\beta\mu),
\end{equation}
where $\beta\equiv 1/(k_{\rm B} T)$, which is precisely known from the MIT measurement in Ref.\,\cite{Ku2012}. Our absorption images indeed give direct access to the pressure of the gas $P$, via the relation \cite{Cheng:2007,Ho:2009,Nascimbene2010,Ku2012,Zwierlein2013book}: 
\begin{equation}
	P(\mu,T) = \frac{m\omega_{x}\omega_{y}}{2\pi }\bar{\bar{n}}(z),
\end{equation}
where $z$ is the axial coordinate, $x,y$ the radial ones, and $\bar{\bar{n}}(z)=\int dx\int dy\,n(x,y,z)$ the doubly integrated density profile, with one of the integrations already provided by the imaging. Within the local density approximation, the chemical potential is given by $\mu = \mu_0 -m\omega_z^2 z^2/2$, with $\mu_0$ the chemical potential at the center of the trap. By fitting the doubly integrated density profile of the gas to the EoS, we are able to extract its temperature $T$.

Key to this approach is the accurate knowledge of the variation in the local chemical potential, which requires a precise knowledge of the trapping potential. This is the motivation for the use of a single ODT at the final stage of evaporation, which, in combination with the magnetic curvature providing trapping in the axial direction, allows us to create a clean and well-calibrated trapping potential. We show in Fig. \ref{Fig:Evap} \emph{in situ} absorption images of our sample at different steps of the evaporative cooling, and the corresponding values of $T/T_{\rm F}$ at the trap center. At the end of evaporation we obtain a cloud at 17(1)\,nK with reduced temperatures of $T/T_{\rm F}=0.08(2)$, well below the critical temperature of the normal-to-superfluid transition $T_{\rm c}/T_{\rm F}=0.176$\,\cite{Ku2012}. Our samples are deliberately prepared at low densities, and we typically obtain a peak average interparticle distance of $n^{-1/3}\simeq1.3\,\mu$m, which makes them compatible with direct imaging via quantum gas microscopy, as this distance is twice larger than the 700\,nm lattice spacing (see Sec.~\ref{sec:lattice}), and is well resolved by our imaging system \cite{ToBePub2023}.

\begin{figure}[t!]
	\centerline{
	\includegraphics[width=\columnwidth]{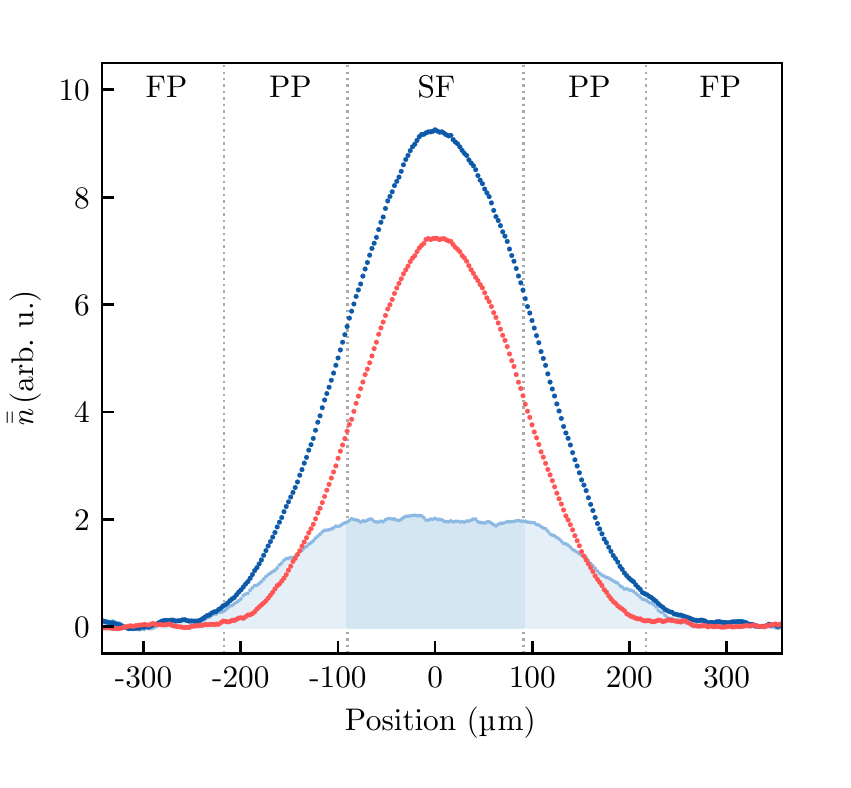}%
	}
	\caption{Superfluid plateau in a spin-imbalanced unitary Fermi gas. Doubly-integrated density profiles for the majority (blue) and minority (red) spin components are shown, obtained for a global spin imbalance of $P\approx 0.25$ and at an estimated temperature of $T\sim 20\,$nK (averaged over 40 experimental realizations). The difference between the two profiles (light-blue points) shows a plateau in the central region, indicating a superfluid core. Vertical lines denote the separation between the superfluid (SF), partially polarized (PP), and the fully polarized (FP) regions.}	
	\label{Fig:Plateau}
\end{figure}

A further, independent proof of superfluidity can be obtained by observing the so-called superfluid plateau, resulting from a phase separation that occurs when spin populations are imbalanced \cite{Shin2006,Zwierlein2006,Shin2008}. Indeed, for a spin-population imbalance below the Clogston-Chandrasekhar limit \cite{Zwerger2012}, a harmonically trapped unitary Fermi gas with a majority of $|1\rangle$, will phase-seperate into three regions: A superfluid core of equal spin densities ($n_1=n_2$), a partially polarized (PP) phase with ($n_1>n_2$) at intermediate distance from the trap center, and a fully polarized (FP) phase ($n_1\neq0$ and $n_2=0$) on the outer part of the trap \cite{Shin2006,Zwierlein2006,Shin2008}. In an axially symmetric trap, the superfluid core can be revealed as a plateau in the difference of the doubly-integrated density profiles of the two components \cite{Shin2006,Shin2008}. 

We prepare a unitary Fermi gas with imbalanced spin-population by adjusting the duration of the RF sweep that takes place before evaporative cooling (see Sec.~\ref{sec:dipole}). As above, the final stage of the evaporation takes place in the single ODT, which is lowered to a trap depth of $\sim k_B \times 200$\,nK. \textit{In situ} density profiles are then recorded through absorption imaging of both spin states concurrently through double exposure of the imaging camera. Fig. \ref{Fig:Plateau} shows doubly integrated density profiles for the majority ($\ket{1}$) and minority ($\ket{2}$) components alongside their difference, which displays a marked central plateau with an axial extent of 200\,$\mu$m, signalling the presence of a superfluid core. This is consistent with our quantitative analysis, with recorded atoms numbers $N_1 = 2.0(2)\times 10^4$ and $N_2 = 1.2(1)\times 10^4$ for the majority and minority components respectively, giving a global spin imbalance $P = \frac{N_1-N_2}{N_1+N_2}\approx0.25$, which is well below the Clogston-Chandrasekhar limit $P_c\sim0.75$ \cite{Shin2006,Zwierlein2006,Lobo2006,Shin2008,Navon2010,Zwierlein2013book}. This demonstrates the ability offered by our apparatus to study the rich physics of strongly-interacting spin-imblanced Fermi gases at ultralow temperatures~\cite{Zwerger2012,Zwierlein2013book}.

\section{Optical Lattices}
\label{sec:lattice}

Our experiment also allows the loading of ultracold samples into optical lattices of adjustable triangular and honeycomb geometry (see Tab.~\ref{Tab:Lattices}). The laser beam configuration used to generate these lattices is shown in the inset of Fig. \ref{Fig:paper_figure}. 

\begin{figure*}[htbp]
	\centerline{
		\includegraphics[width=\textwidth]{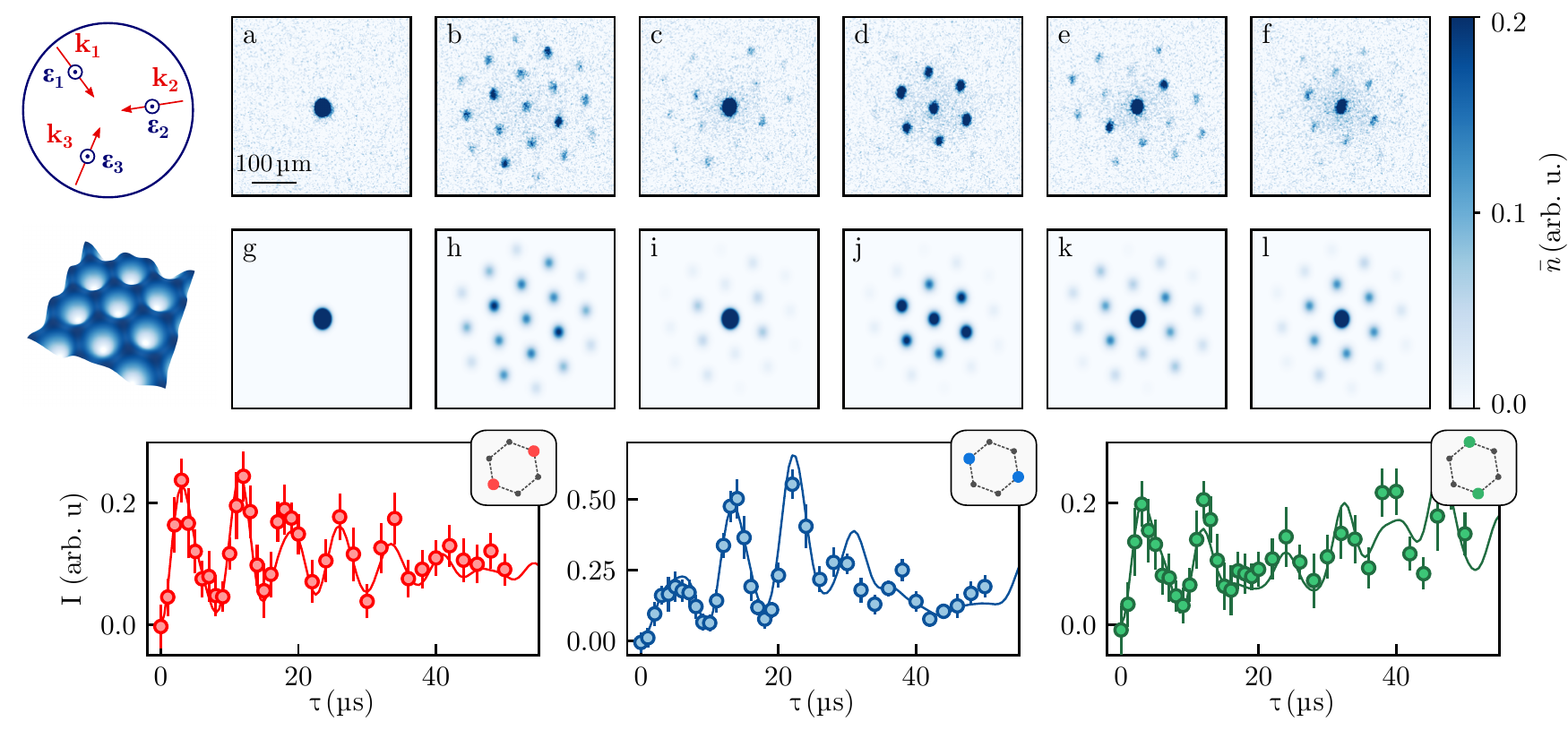}
	}
	\caption{Bragg scattering of a molecular BEC exposed to an optical lattice with triangular geometry. The configuration of the wave vectors $\boldsymbol{k_i}$ and polarization vectors $\boldsymbol{\epsilon_i}$ used to generate the lattice is shown in the left inset, alongside a schematic depiction of the lattice geometry. Experimental (top row) and simulated (bottom row) absorption images following a $1.5\,\mathrm{ms}$ TOF after exposing the cloud to the lattice for a pulse duration $\tau$ of (a, g) 0; (b, h) 5; (c, i) 9; (d, j) 12; (e, k) 18; (f, l) 132\,$\mu$s. The bottom three panels show the population ($I$) as a function of $\tau$ for the six first order diffraction peaks. Corresponding peaks for each graph are indicated by the top-right insets. Experimental data points are given with error bars representing one standard deviation together with results from the simulation (solid lines). Data is averaged over 15 experimental realizations.}
	\label{Fig:BraggT}
\end{figure*}

\begin{figure*}[htbp]
	\centerline{
		\includegraphics[width=\textwidth]{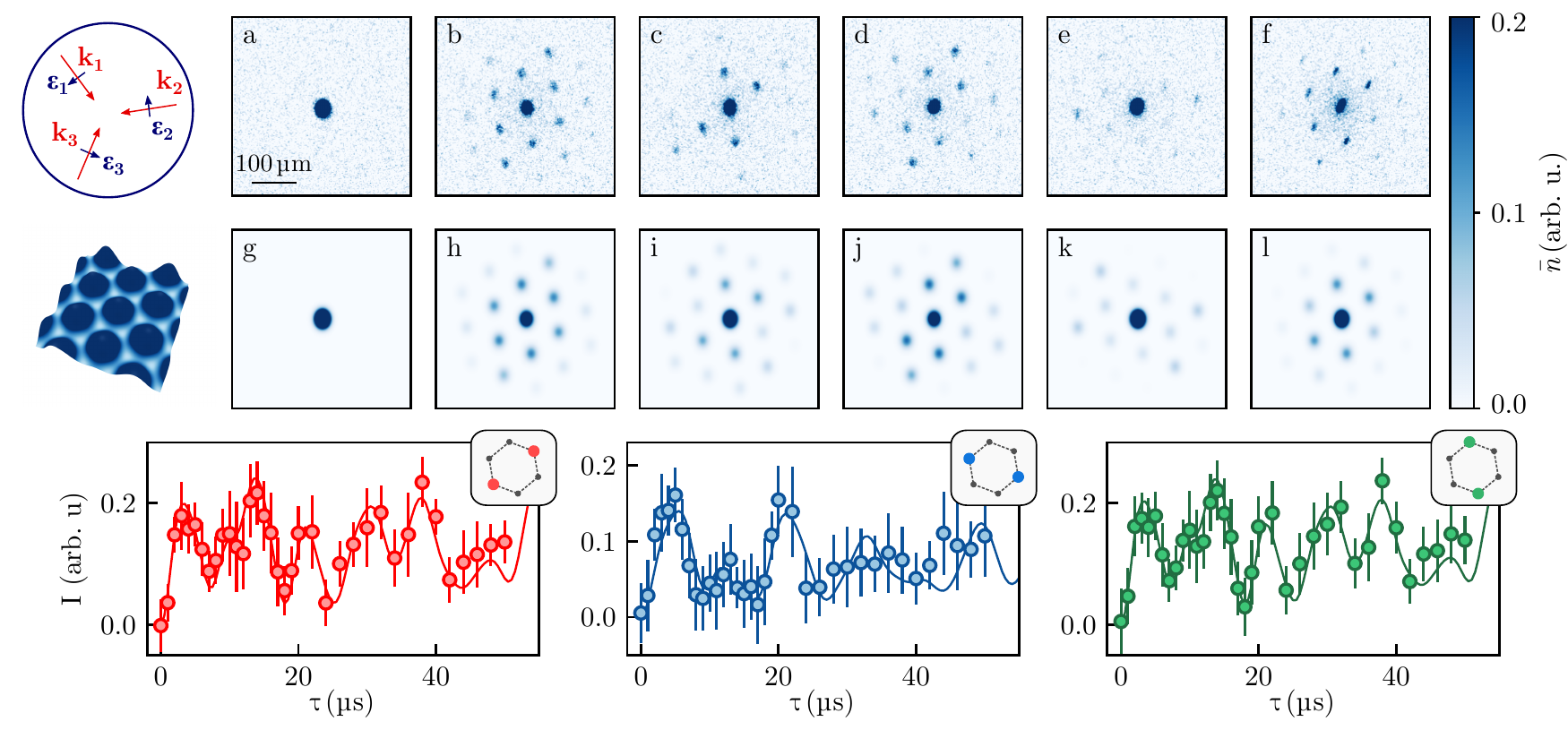}
	}
	\caption{Bragg scattering of a molecular BEC exposed to an optical lattice with  honeycomb geometry. The configuration of the wave vectors $\boldsymbol{k_i}$ and polarization vectors $\boldsymbol{\epsilon_i}$ used to generate the lattice is shown in the left inset, alongside a schematic depiction of the lattice geometry. Absorption images taken following a $1.5\,\mathrm{ms}$ TOF after exposing the cloud to the lattice for a pulse duration $\tau$ of (a, g) 0; (b, h) 5; (c, i) 9; (d, j) 12; (e, k) 18; (f, l) 132\,$\mu$s. The bottom three panels show the population ($I$) as a function of $\tau$ for the six first order diffraction peaks. Corresponding peaks for each graph are indicated by the top-right insets. Experimental data points are given with error bars representing one standard deviation together with results from the simulation (solid lines). Data is averaged over 15 experimental realizations.}
	\label{Fig:BraggH}
\end{figure*}

In the horizontal (XY) plane, a single laser beam in butterfly configuration forms three arms with relative angles of approximately $120^\circ$. Interference of the three arms can be used to generate either a triangular or a honeycomb lattice geometry by an appropriate choice of the arms' polarization, which is controlled by rotating a single half-waveplate. For the triangular lattice, the polarization vectors of the three lattice arms are parallel and lie along the vertical (Z) direction, while for a honeycomb lattice the polarization vectors lie in the XY-plane \cite{Grynberg:1993,Becker2010}. The specific wave vector and polarization vector configurations are schematically illustrated in the insets of Figures \,\ref{Fig:BraggT} and \ref{Fig:BraggH}. The same laser beams can also be used in a non-interfering configuration to form a deep CDT merely by rotating the aforementioned waveplate to set the polarization of the incident beam at an angle of $\mathrm{arctan}(\sqrt{2}) \simeq 55^{\circ}$ with respect to the Z-axis, such that the polarization vectors of all lattice arms are mutually orthogonal.

\begin{table}[]
\centering
\caption{Different potential configurations available in the science cell. The horizontal plane (XY) can host a triangular (T) or honeycomb (H) lattice (latt.), whereas in the vertical direction (Z) we can confine the particles in a one-dimensional lattice, a light sheet or a crossed optical dipole trap.}
\label{Tab:Lattices}
\begin{tabular}{c | c c c}
\hline
\hline
\backslashbox{\textbf{XY \hspace{14 pt}}}{\textbf{\hspace{14 pt} Z}}&  \textbf{Z lattice}  & \textbf{Light sheet}  & \textbf{Z CDT}  \\ \hline
\textbf{Off}         & 2D Layers        & 2D Gas         & 3D Gas        \\ 
\textbf{Triangular}  & Layered T latt.  & 2D T latt.     & 1D T array  \\ 
\textbf{Honeycomb}   & Layered H latt.  & 2D H latt.     & 1D H array  \\ 
\hline
\hline
\end{tabular}
\end{table} 

In the Z-direction, a one-dimensional lattice is created with a pair of laser beams crossing at $90^\circ$. The strength of the interference between the two beams is tuned via an electronically controlled half-waveplate which allows for in-sequence ramping from a CDT to a one-dimensional lattice. To complement the vertical lattice, which allows us to produce a stack of 2D systems, our experiment also features the ability to produce a single layer of atoms using a highly oblate laser beam, or light sheet, propagating along the Y-direction and providing a strong vertical confinement of $65$\,kHz. Beams for the XY- and Z-lattices, as well as the light sheet are created with three independent laser beams at 1064 nm, with a maximum power of 40\,W for the former two and 16\,W for the latter. 

The triangular or honeycomb lattices in the XY plane can therefore be combined with either the vertical lattice, allowing us to create a stack of layers with tunable interlayer coupling, or with the single light sheet, giving access to purely 2D physics that can be readily probed by single-atom imaging \cite{ToBePub2023}. In the following, we validate the multifunctionality of our apparatus by a quantitative characterization of the two XY-lattices.

\begin{figure}[htbp]
	\centerline{
		\includegraphics[width=\columnwidth]{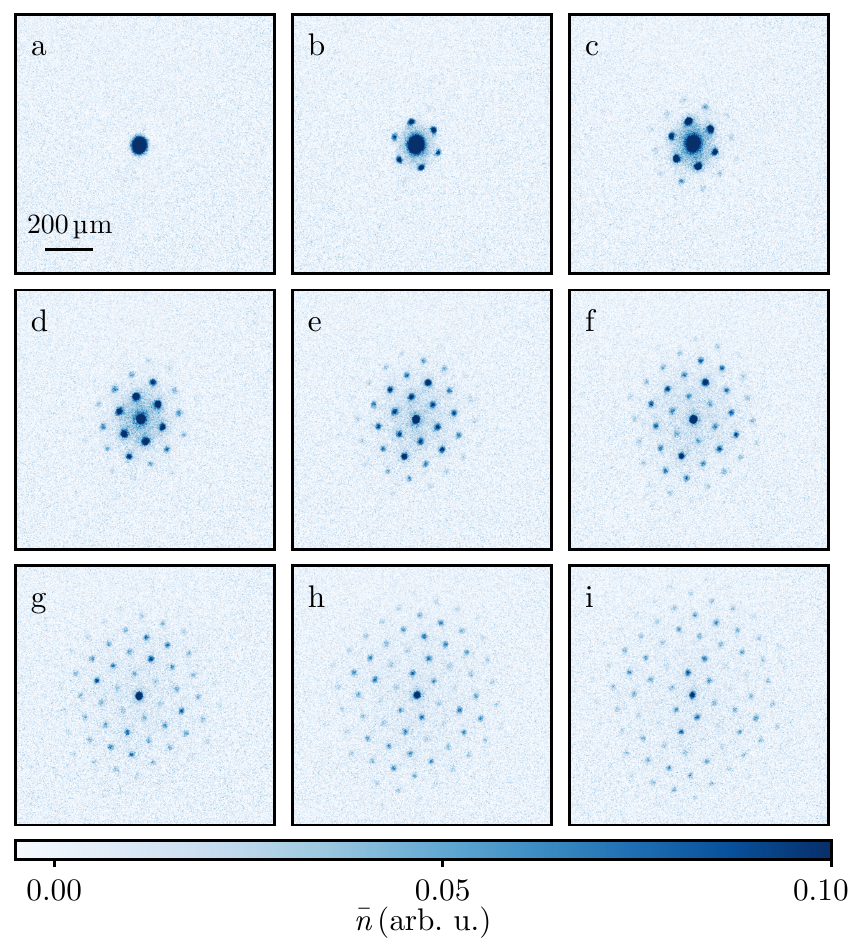}
	}
	\caption{Absorption images showing Kapitza-Dirac diffraction of a molecular BEC taken after a 2\,ms time-of-flight for different values of the geometric average of the pulse area $\bar{\theta}$. Images for (a) $\bar{\theta}=0$, (b) $\bar{\theta} = 0.49$, (c) $\bar{\theta} = 0.73$, (d) $\bar{\theta} = 1.08$, (e) $\bar{\theta} = 1.43$, (f) $\bar{\theta} = 1.74$, (g) $\bar{\theta} = 2.05$, (h) $\bar{\theta} = 2.43$ and (i) $\bar{\theta} = 2.77$ averaged over 60 experimental realizations. See Appendix \ref{App:KD} for the corresponding pulse times and potential depths.}
	\label{Fig:KD}
\end{figure}

We first characterize the lattices using Kapitza-Dirac scattering \cite{Kapitza1933,Gupta2001}. This technique consists of pulsing the lattice potential on a matter wave and subsequently performing a TOF expansion, yielding a diffraction pattern that reflects the momenta imparted by the lattice, which provides a robust calibration of the lattice depth and axes orientations. Indeed, upon exposure to a lattice with trap depth $U_{0}$ for a duration $\tau$, the particles are distributed over several momentum classes whose populations solely depend on the pulse area $\theta = U_{0} \tau/(2\hbar)$, under the condition that the particle motion can be neglected during the pulse time (Raman-Nath approximation) \cite{Gupta2001}. This result can be readily generalized to the case where the lattice arms have different intensities. For ultracold systems, Kapitza-Dirac scattering has been first observed for BECs (of bosonic atoms) \cite{Ovchinnikov1999}, and more recently, strongly-interacting molecular BECs were used to investigate the role of interactions in the scattering from a one-dimensional lattice \cite{Liang2022}.

For this measurement, instead of using the CDT described in Sec.\,\ref{sec:evap}, we load the atom cloud in the CDT created by the laser beams of the vertical lattice with their polarizations set to be orthogonal. There we perform two successive evaporation ramps of 1\,s each, first at 832\,G and then at 740\,G to ensure the formation of a large molecular BEC. After that, we ramp the magnetic field deeper into the BEC side, down to 665\,G (10\,ms ramp down followed by a 10\,ms holding time). At this magnetic field the intermolecular scattering length is 600\,$a_0$, allowing us to neglect the effects of interparticle interactions during TOF expansion given our densities ($na^3<10^{-4}$). To perform the scattering experiment, we first release the molecular BEC from the CDT for $\lesssim1\,\mu$s and then shine the triangular XY-lattice on the molecular BEC for a pulse time $\tau$ ranging from 500\,ns to 1\,$\mu$s with optical power ranging between 0 and $4\,$W. We then perform a brief TOF of 2\,ms after which we take absorption images. Typical diffraction patterns for various pulse areas are shown in Fig.\,\ref{Fig:KD}.

From the diffraction pattern structure we deduce the beam intersection angles with respect to the camera horizontal axis $\phi_1, \phi_2, \phi_3$ and the imbalance between the lattice arm intensities. We find $\phi_1 = 190^\circ$, $\phi_2 = 311^\circ$, $\phi_3 = 72^\circ$; a beam imbalance ratio of $0.74:0.85:1.4$ and a maximum achievable lattice depth of $888(24)\,\mu$K. We performed similar measurements to characterize the one-dimensional vertical lattice, as shown in Appendix\,\ref{App:KD}, resulting in a maximum lattice depth of  $401(13)\,\mu$K and trap frequencies up to 701(11)\,kHz in the vertical direction, allowing for wide tunability for the confinement and tunneling in the vertical direction.

While the Kapitza-Dirac measurement is a reliable method to calibrate the lattice depth and identify the lattice axes, we find that it does not distinguish between the triangular and honeycomb configurations, as it yields the same diffraction patterns in both cases, within a global factor. To understand this observation, we have developed a general scattering model that we present in Appendix~\ref{App:BraggScattering}, together with a detailed theoretical analysis. Using the full Hamiltonian of the problem, we find that lattice pulse times at least on the order of the lattice trap period -- hence beyond the Kapitza-Dirac regime -- are required to identify the triangular and honeycomb geometries unambiguously. For these longer pulse times, we enter the Bragg scattering regime \cite{Bragg1913} where molecules from the condensate are transferred to a set of discrete momentum states corresponding to wave vectors of the reciprocal lattice. The dynamics at play can then be viewed as resulting either from coherent two-photon processes involving the modes making up the standing waves of the lattice \cite{Gupta2001}, or from the diabatic projection of the free-space momentum eigenstates onto those of the lattice \cite{Denschlag2002}. In the absence of decoherence processes, the dynamics only depends on the lattice recoil energy $E_\mathrm{L}$, and the two-photon Rabi frequencies $\Omega_{12}$, $\Omega_{13}$ and $\Omega_{23}$ resulting from each pair of lattice arms.

The predicted difference in the diffraction dynamics mainly originates from the sign of the off-diagonal terms in the full Hamiltonian, which are exactly opposite for the triangular and honeycomb lattices. This sign difference has significant consequences as it tunes the two-photon process closer to or further from resonance, with marked effects on the momentum populations at sufficiently long time-scales.

We perform Bragg scattering with lattice pulse times $\tau$ up to $150\,\mu$s, and observe rich dynamics with striking differences between the two configurations, which we unambiguously differentiate and identify by comparison with our theoretical predictions. Absorption images taken after a $1.5$\,ms TOF are shown in Fig.\,\ref{Fig:BraggT} (Fig.\,\ref{Fig:BraggH}) for the triangular (honeycomb) configuration, where we used a lattice power of $200\,$mW ($400\,$mW), such that $\hbar |\Omega_{12} | \simeq\hbar |\Omega_{13} | \simeq\hbar |\Omega_{23} | \simeq E_\mathrm{L}$. Quantitatively, we also display the time evolution of the populations of the different diffraction orders in the two lattice geometries, which are well reproduced by our model, using the intensity of each lattice arm and a relaxation coefficient as fitting parameters. The resulting traces are displayed in the bottom panels of Figs. \ref{Fig:BraggT} and \ref{Fig:BraggH}, and show excellent agreement with the predictions for both geometries. Bragg scattering thus provides a sensitive way to characterize the trapping frequencies in different directions of optical lattices even when those are nearly degenerate, which is a regime where traditional methods such as amplitude/phase modulation or Raman spectroscopy are less resolved due to thermal or Doppler broadening.

\section{Conclusions}

We have introduced a new platform for quantum simulation experiments based on $^6$Li atoms. Our apparatus features two chambers, with a laser-cooling chamber and a science cell with excellent optical access. We combine the use of $D_1$ gray molasses sub-Doppler cooling with the loading into a mechanically movable optical dipole trap, which proves to be robust and allows us to benefit from this two-chamber architecture without major technical overhead. Thanks to an all-optical cooling strategy that only relies on 671\,nm and near-infrared laser wavelengths, we reliably produce deeply degenerate Fermi gases in approximately 15\,s cycles. A key characteristic of our apparatus is its versatility, which allows placing the degenerate samples in a variety of energy landscapes without any hardware change. 

We created unitary Fermi gases (in 3D continuous space) at controlled temperatures, which we obtain using \textit{in situ} thermometry. With temperatures as low as $0.08\,T_{\rm F}$, we prepared superfluid samples in a regime of high diluteness corresponding to a peak interparticle spacing larger than a micron, bringing their direct imaging via quantum gas microscopy within reach \cite{ToBePub2023}. 

We generated a versatile set of optical lattices enabling the study of interacting fermionic matter in continuous 2D and 1D space, as well as in multiple lattice configurations. In the horizontal plane, we showed the tunability from triangular to honeycomb lattice, which we characterized via matter-wave diffraction of a molecular Bose-Einstein condensate, in and beyond the Kapitza-Dirac regime. We demonstrated how Bragg diffraction can be used to quantitatively discriminate between the two and overcome the limitations of standard Kapitza-Dirac diffraction. In the vertical direction, trapping can be provided by a one-dimensional lattice, allowing us to create a stack of layers with tunable interlayer coupling, or with a single light sheet, giving access to purely 2D physics. 

The same lattices, in combination with the high-resolution objective and Raman sideband cooling, also enable single-atom imaging in this apparatus, which is already operational and is the subject of ongoing work \cite{ToBePub2023}. Single-atom imaging of dilute deeply degenerate Fermi gases offers the prospect to directly access spin-resolved spatial correlation functions of the unitary Fermi gas, which were never measured to date and would provide a unique microscopic characterization of this paradigmatic many-body system. Such observables are particularly crucial in the regime of spin-imbalance to probe the physics of interacting Fermi polarons \cite{Muir2022} or search for the elusive FFLO phase \cite{Fulde1964,Larkin1964} that was predicted 60 years ago but never observed. Our setup also offers the perspective to probe the microscopics of spin systems in triangular and honeycomb lattices, where the interplay between magnetic correlations and frustration is expected to give rise to a rich and highly debated phenomenology, including possible spin liquid phases \cite{Shimizu2003,Meng2010,Sorella2012,Assaad2013,Gu2013,Ferrari2017,Shirakawa2017,Zhang2019,Li2020,Szasz2020,Tocchio2020,Wietek2021}.

With its extended experimental capabilities, the quantum gas platform presented here enables a new approach for atom-based quantum simulation, directed towards addressing different classes of quantum many-body systems within the same setup.\\

\begin{acknowledgments}
We thank Clara Bachorz, Darby Bates, Markus Bohlen, Valentin Cr\'epel, Yann Kiefer, Joanna Lis, Mihail Rabinovic, Julian Struck for experimental assistance in the early stages of this project, and Sebastian Will for a critical reading of the manuscript. This work has been supported by Agence Nationale de la Recherche (Grant No. ANR-21-CE30-0021), the European Research Council (Grant No. ERC-2016-ADG-743159), CNRS (Tremplin@INP 2020), and R{\'e}gion Ile-de-France in the framework of DIM SIRTEQ (Super2D and SISCo) and DIM QuanTiP.

\end{acknowledgments}

\appendix

\section{Experimental apparatus and Sequence}
\label{App:MOT}
Here we present further information on the experimental apparatus and sequence.

Table\,\ref{tab:laser} summarizes the typical laser beam parameters of the laser cooling stages. The MOT laser beams, operating near the $D_2$ transition ${(^2S_{1/2} \rightarrow {^2P_{3/2})}}$, consist of two retro-reflected beams and one counter-propagating beam pair, all with a $1/e^2$ radius of 7.5\,mm. After the MOT stage we perform a compression (CMOT) by increasing the magnetic field gradient from $10$\,G/cm to $25$\,G/cm over a period of 50\,ms. We then suddenly switch off the magnetic field gradients, while keeping the MOT laser beams on until all transient magnetic gradients have decayed (we refer to this as the $D_2$ molasses phase, despite the presence of transient magnetic fields). We subsequently capture the atoms in a gray molasses, based on the $D_1$ transition ${(^2S_{1/2} \rightarrow {^2P_{1/2})}}$, for which we use two retro-reflected beams with a $1/e^2$ radius of 3\,mm and one retro-reflected beam --- overlapped with one of the MOT arms --- with a $1/e^2$ radius of 7.5\,mm. Atoms are then optically pumped into the $F=1/2$ manifold by switching off the weak $D_1$ molasses beams for 10$\,\mu$s and moving the strong beam frequency towards resonance. This ensures that the atoms populate the two Zeeman sublevels used for evaporative cooling.

\begin{table}[t]
  \centering
  \caption{Typical experimental parameters for the laser beams involved in the laser cooling stages. We refer to the two beams involved in the $D_1$ molasses stage as strong and weak, respectively, as each contributes to cooling due to the $\Lambda$-enhancement of the gray molasses \cite{Grier2013, Sievers2015, Dash2022}. Angular momentum of the ground ($F_g$) state and addressed transitions are given for each cooling step, alongside detunings ($\delta$) expressed in units of the natural linewidth $\Gamma = 2\pi\times 5.87\,$MHz, laser intensities per arm ($I$) in units of the saturation intensity $I_\mathrm{sat} = 2.54$ mW/cm$^2$ of the $D_2$ transition of $^6$Li, as well as the total optical power used per cooling stage $(P_\mathrm{tot})$.}
\begin{tabular}{lccccc}
\hline\hline
Laser Beam         &$F_g$ & Transition& $\delta$     & $I$        & $P_\mathrm{tot}$ \\
                   &      &           & ($\,\Gamma$) & ($I_\mathrm{sat}$) & (mW)\\ \hline
(C-)MOT  Cooling   & $3/2$& $D_2$     &-3.4          &  3.6       & 35  \\
(C-)MOT  Repumper  & $1/2$& $D_2$     &-3.4          &  1.3       & 13  \\
$D_2$ mol. Cooling  & $3/2$& $D_2$     &-1            &  0.2       & 4   \\
$D_2$ mol. Repumper    & $1/2$& $D_2$     &-1            &  0.1       & 1   \\
$D_1$ mol. Strong  & $3/2$& $D_1$     & 4            &  $\sim 20$     & 60  \\
$D_1$ mol. Weak    & $1/2$& $D_1$     & 4            &  $\sim 1$      & 2   \\ \hline\hline
\end{tabular}
  \label{tab:laser}
\end{table}

We then capture the atoms in the ODT (ODT capture stage) and transport them to the glass cell. To estimate the transport efficiency, we measure the atom number after a round-trip transport, i.e., to the glass cell and back to the MOT chamber. This allows us to count the atom number under identical conditions as before the transport. The efficiency $\epsilon$ is then obtained by computing:
\begin{equation*}
\epsilon = \sqrt{\frac{N_\mathrm{RT}}{N(\Delta t)}}
\end{equation*}
where $N_\mathrm{RT}$ is the atom number after the round trip and $N(\Delta t)$ is the atom number after a hold time of $\Delta t = 2.4\,$s in the MOT chamber, corresponding to the duration of the round trip. We obtain a transport efficiency of 97\% with a standard error of $0.5\%$.

The atoms are subsequently loaded in the CDT and the magnetic field is ramped to 832\,G. The RF sweep is then performed to equilibrate the population (ODT RF), and evaporation is initiated by lowering the power of the transport beam until it becomes comparable to the power of the crossing beam (CDT balanced). Typical atom numbers, temperatures and phase-space densities obtained at the end of each of these cooling stages are presented in Table\,\ref{tab:atom}.

In the glass cell, atoms are detected via absorption imaging along either a horizontal or vertical axis. The majority of the data presented here is obtained using the vertical imaging system with a magnification of $\sim 6$, which consists of a microscope objective corrected to infinity followed by a first focusing lens and a telescope projecting the image of the atoms on an EMCCD camera (Andor iXon Ultra 888). The lens setup is designed to easily switch to a magnification of $\sim 60$, which allows for single atom imaging and the best performance of the microscope objective (not used here). The latter features a numerical aperture of 0.56, an effective focal length of 27\,mm, with optimal resolution $\lesssim 1 \,\mu$m over a field-of-view of $200\,\mu$m. The imaging setup in the horizontal plane consists of two lenses in a 4$f$-configuration and a CMOS camera (Andor Zyla 5.5), resulting in a magnification of 1.5.

\begin{table}[t]
  \centering
  \caption{Typical atom numbers ($N$), temperatures ($T$) and phase space densities (PSD) at the end of each stage of the cooling sequence up to the end of the first evaporation step.}
  \begin{tabular}{llll}
\hline\hline
               & N             & T           & PSD                \\ \hline
CMOT           & $1\times10^9$ & $1.2$\,mK   & $5.5\times10^{-7}$ \\
$D_2$ molasses & $1\times10^9$ & $800\,\mu$K & $5.8\times10^{-7}$ \\
$D_1$ molasses & $5\times10^8$ & $50\,\mu$K  & $5.2\times10^{-5}$ \\
ODT capture    & $5\times10^6$ & $90\,\mu$K  & $3.4\times10^{-4}$ \\
ODT RF         & $2\times10^6$ per spin & $125\,\mu$K & $5.0\times10^{-5}$ \\
CDT balanced          & $3.4\times10^5$ per spin & $23\,\mu$K  & $2.0\times10^{-2}$ \\\hline\hline
\end{tabular}
  \label{tab:atom}
\end{table}

\section{Kapitza Dirac Scattering}
\label{App:KD}
Here we provide additional experimental parameters for the Kapitza-Dirac measurements presented in the main text, performed for the two-dimensional lattice triangular configuration. We also present the Kapitza-Dirac measurements used to characterize the one-dimensional vertical lattice.

Table\,\ref{tab:KDHT} shows the experimental parameters of the two-dimensional lattice that were used for the Kapitza-Dirac measurements shown in Fig.\,\ref{Fig:KD}. Resulting momentum state populations only depend on the pulse area $\theta = U_0\tau/(2\hbar)$.

\begin{table}[b]
  \centering
  \caption{Pulse area $\theta$, pulse time $\tau$ in $\mu$s and potential depth $U_0/k_B$ in $\mu$K for each of the Kapitza-Dirac measurements shown in Fig\,\ref{Fig:KD}. Labels a through i refer to the corresponding panels.}
\begin{tabularx}{0.975\columnwidth}{l *{9}{Y}}
\toprule
 & \multicolumn{1}{c}{a} & \multicolumn{1}{c}{b} & \multicolumn{1}{c}{c} & \multicolumn{1}{c}{d} & \multicolumn{1}{c}{e} & \multicolumn{1}{c}{f} & \multicolumn{1}{c}{g} & \multicolumn{1}{c}{h} & \multicolumn{1}{c}{i} \\ \hline
$\theta$  & 0.0 & 0.49 & 0.73 & 1.08 & 1.43 & 1.74 & 2.05 & 2.44 & 2.77 \\
$\tau$    & 0.0 & 1.00 & 1.00 & 0.80 & 0.80 & 0.60 & 0.60 & 0.50 & 0.50 \\
$U_0/k_B$ & 0.0 & 3.8  & 5.5  & 10.3 & 13.6 & 22.2 & 26.3 & 37.0 & 42.1 \\
\hline\hline
\end{tabularx}
  \label{tab:KDHT}
\end{table}

\begin{figure}[b!]
	\centering
	\includegraphics[width=\columnwidth]{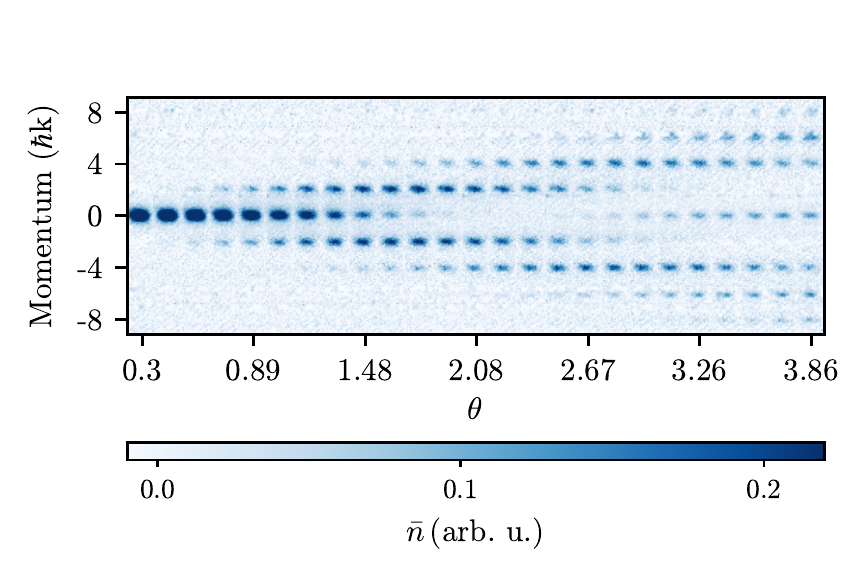}
	\caption{Kapitza-Dirac measurement in the one-dimensional vertical lattice with $\tau = 450$\,ns. Each column shows an absorption image taken at a specific laser power, varying from 0.5 to 6.5\,W. Values of the pulse area $\theta$ are obtained by fitting the population of the momentum states to the respective Bessel functions, as shown in Fig.\,\ref{Fig:KDzBessel}.}
	\label{Fig:KDz}
\end{figure}

Similar to the measurements performed in the two-dimensional lattice, we perform Kapitza-Dirac scattering in the one-dimensional, vertical lattice by preparing a molecular BEC in the CDT at a magnetic field of 665\,G. We then turn off the CDT and pulse the lattice for a time $\tau = 450$\,ns at a varying laser power of the lattice beams $P$ between 0.5 and 6.5\,W, leading to pulse areas $\theta$ between $0.3$ and $3.86$. After a brief TOF of $1.8$\,ms we capture the resulting diffraction patterns through absorption imaging along the horizontal imaging axis, which we show in Fig.\,\ref{Fig:KDz}.

The simple geometry of the vertical lattice allows us to describe these diffraction patterns with Bessel functions. Following Refs.\,\cite{Gupta2001,Ovchinnikov1999} we write the light-shift potential of the far-detuned lattice beams onto the molecules along the vertical ($\mathbf{z}$) direction as:
\begin{equation*}
U(z,t) = U_0 f^2(t)\sin^2(\mathbf{k\cdot z}),
\end{equation*}
where $f(t)$ is the temporal envelope of the lattice beams, in our case a block pulse of time $\tau$, and $\mathbf{k}$ is the lattice wave vector. In the Raman-Nath regime, where we can ignore kinetic energy contributions, we can use simple time evolution to find the wave function $\ket{\psi(\tau)}$ from the initial (zero-momentum) wave function $\ket{\mathbf{p=0}}$:
\begin{equation*}
\ket{\psi(\tau)} = \exp\left(-\frac{i}{\hbar}\int_0^\tau dt\, U(z,t)\right)\ket{\mathbf{p}=\mathbf{0}}.
\end{equation*}
The result can be rewritten in a sum of plane waves, i.e., momentum eigenstates $\ket{\mathbf{p}=2n\hbar \mathbf{k}}$ with $n\in\mathbb{Z}$, with $n^\mathrm{th}$ order Bessel functions ($J_n$) as pre-factors. This leads to populations in the respective momentum states of:
\begin{equation*}
P_n(\theta) = J^2_n(\theta) \;\; \mathrm{with}\; \theta = \frac{U_0\tau}{2\hbar}.
\end{equation*}

\begin{figure}
	\centerline{
		\includegraphics[width=\columnwidth]{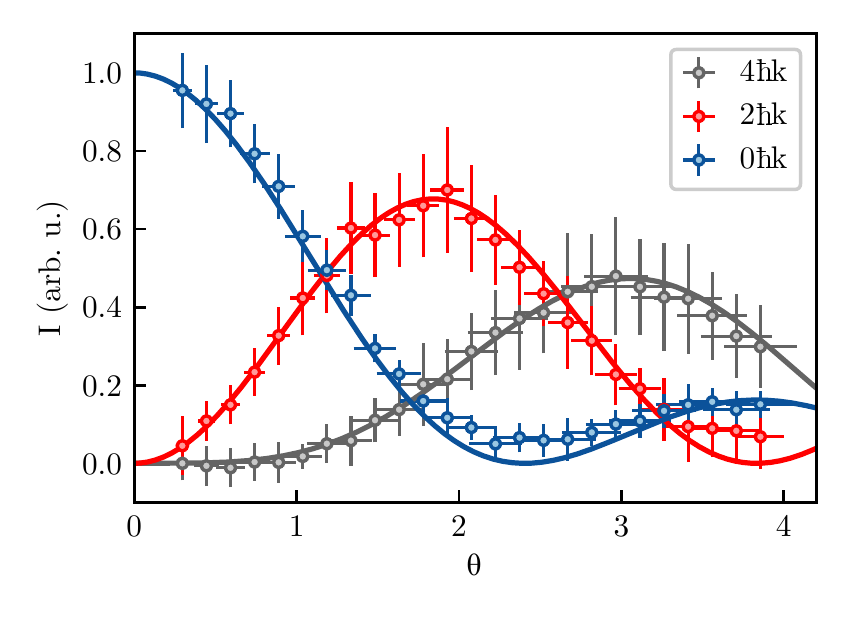}
	}
	\caption{Population in each momentum eigenstate $(I)$ for the Kapitza-Dirac measurement shown in Fig.\,\ref{Fig:KDz} as a function of pulse area $(\theta)$. We scale the horizontal axis to fit the experimental results (data points) with the expected Bessel function behavior (solid lines). Vertical error bars show experimental fluctuations at one standard deviation, while horizontal error bars represent the uncertainty of the fit at one standard deviation.}
	\label{Fig:KDzBessel}
\end{figure}

We determine the population in each momentum state and for each lattice beam power for the measurements shown in Fig.\,\ref{Fig:KDz}. By fitting the results to the Bessel functions for the respective momentum states, we calibrate the trap depth of the vertical lattice as a function of the optical power of the lattice beam $P_z$. Note that we sum the population of the eigenstates with the same absolute momentum. Resulting data points and fits for the first three momentum states are shown in Fig.\,\ref{Fig:KDzBessel}, leading to an approximate atomic trap depth power dependency of $k_B \times 10$\,$\mu$K\,W$^{-1}$\,$\times P_z$. We used a similar approach to analyze the Kapitza-Dirac scattering results for the XY-lattice presented in the main text (Fig.\,\ref{Fig:KD}).

\section{Model of BEC Scattering off a Periodic Potential}
\label{App:BraggScattering}

\begin{figure*}[h!t]
	\centerline{
		\includegraphics[width=\textwidth]{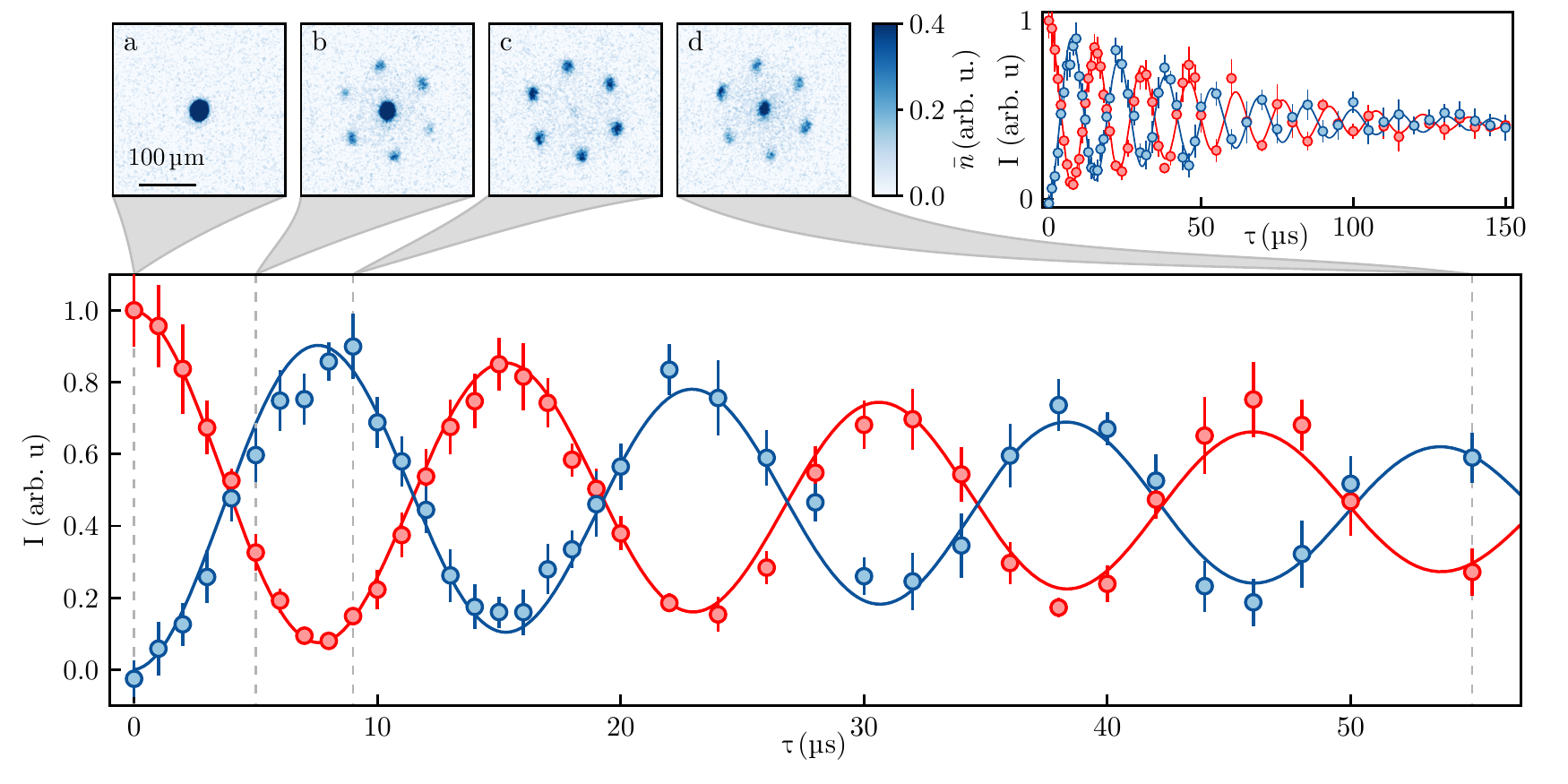}
	}
	\caption{Bragg scattering of a molecular BEC exposed to a triangular lattice with $P \simeq 100\,\mathrm{mW}$, corresponding to a two-photon Rabi frequency $|\Omega| \simeq 2 \pi \times 28(2)\,\mathrm{kHz}$. Absorption images taken after 1.5\,ms TOF for $\tau =$ (a) 0 , (b) 5, (c) 8, and (d) 55\,$\mu$s. The main panel shows the integrated signal for the zeroth order diffraction peak (red data points) and the first order diffraction peaks (sum of the six degenerate first order peaks, blue data points) as a function of lattice pulse duration up to $\tau = 55$\,$\mu$s. Corresponding fits of the damped Rabi oscillation (red and blue lines, respectively) are also shown, yielding an effective Rabi frequency of $2\pi\times 64.3(5)\,\mathrm{kHz}$, close to the $\sqrt{6}\Omega$ value predicted by our two-level model, and a relaxation time of $59(6)\,\mu$s. The top-right inset shows the population over an extended range with $\tau $ up to 150\,$\mu$s.}
	\label{Fig:Bragg}
\end{figure*}

In this section we detail our model of the scattering of a molecular BEC on a periodic potential $V(\mathbf{r})$ that is similar to those presented in Refs.\,\cite{Viebahn2019,Sbroscia2020,Kosch2022}. We apply this model to the Bragg-scattering measurements in the XY-lattice presented in Figs.\,\ref{Fig:BraggT}, \ref{Fig:BraggH} and \ref{Fig:Bragg}, but it is also suitable to describe Kapitza-Dirac scattering. We additionally derive an effective two-level model for a triangular or honeycomb lattice in the case where the intensities of all three lattice beams are balanced and low enough to suppress occupation of large momentum states.

The optical potential of the XY-lattice is created by the interference of three co-planar beams, which we write as:
\begin{equation*}
\label{Eq:optical_potential}
V(\mathbf{r}) = -A \left| E_1 e^{i\mathbf{k}_1\mathbf{\cdot r}} \boldsymbol{\epsilon}_1 +  E_2 e^{i\mathbf{k}_2\mathbf{\cdot r}} \boldsymbol{\epsilon}_2 +  E_3 e^{i\mathbf{k}_3\mathbf{\cdot r}} \boldsymbol{\epsilon}_3\right|^2,
\end{equation*}
where $A$ is a constant, $\mathbf{r}$ the position vector and the $E_i$, $\mathbf{k}_i$ and $\boldsymbol{\epsilon}_i$ are respectively the electric field amplitude, wavevector and polarization vector for each of the three beams. 

Expanding this potential leads to a constant term $V_0$, as well as $6$ interference terms $\mathrm{exp}(i (\mathbf{k}_i - \mathbf{k}_j) \cdot \mathbf{r})$ for $i\neq j$  with potentially different amplitudes depending on the local intensity and polarization  of the beams. These terms can equally be seen as a position-dependent potential, or as a coupling between different momentum states, since in the momentum basis: $\mathrm{exp}(i (\mathbf{k}_i - \mathbf{k}_j) \cdot \mathbf{\hat{r}}) \left|\mathbf{p}\right\rangle = \left|\mathbf{p} + \hbar (\mathbf{k}_i - \mathbf{k}_j)\right\rangle$.

In our geometry, $\mathbf{k}_1$, $\mathbf{k}_2$ and $\mathbf{k}_3$ lay in the XY-plane at $120^{\circ}$ angles with each other. Considering an initial molecular BEC prepared in a single momentum state $\left| \mathbf{p}=\mathbf{0}\right\rangle$, the set of accessible momentum states forms a (reciprocal) triangular Bravais lattice $\left\{\left|n_1 \mathbf{b}_1 + n_2 \mathbf{b}_2\right\rangle, (n_1,n_2) \in \mathbb{Z}^2 \right\}$. Specifically, we denote the six non-zero momentum states surrounding the origin with $\mathbf{b}_1 = \hbar(\mathbf{k}_1-\mathbf{k}_2)$, $\mathbf{b}_2 = \hbar(\mathbf{k}_3-\mathbf{k}_2)$, $\mathbf{b}_3 = \hbar(\mathbf{k}_3-\mathbf{k}_1)$, $\mathbf{b}_4 = - \mathbf{b}_1$, $\mathbf{b}_5 = - \mathbf{b}_2$ and $\mathbf{b}_6 = - \mathbf{b}_3$.

The Hamiltonian for this system is composed of kinetic and potential energy terms $H = T + V$. Writing two momentum eigenstates as $\ket{\alpha}$ and $\ket{\beta}$ this gives:
\
\begin{align*}
&\left\langle \alpha \right| T \left| \beta \right\rangle = \delta_{\alpha,\beta} \frac{\mathbf{b}_{\alpha}^2}{2M} \equiv \delta_{\alpha,\beta} \hbar \Delta_\alpha,\\
&\left\langle \alpha \right| V \left| \beta \right\rangle = \left\{\begin{aligned}
&\frac{\hbar \Omega_{ij}}{2} \;
\text{if}\; \frac{\mathbf{b}_{\alpha}-\mathbf{b}_{\beta}}{\hbar} = \mathbf{k}_i - \mathbf{k}_j, i \neq j \in   [1 ... 3] \\
&0 \quad
\text{otherwise}
\end{aligned} \right.,
\end{align*}
where, by construction $\hbar \Delta_1 = \hbar \Delta_2 = ... = \hbar \Delta_6 = 3\hbar^2 k^2/(2M) = E_\mathrm{L}$; and the Rabi frequencies $\hbar \Omega_{ij} = -2 A E_i E_j \boldsymbol{\epsilon}^*_i \cdot \boldsymbol{\epsilon}_j$ depend on the relative intensity and polarization of the lattice beams. 

By diagonalizing this Hamiltonian, we are able to explore the Bragg dynamics for different parameters, which show that the evolution of the diffraction patterns as a function of the lattice pulse are markedly different between the triangular and the honeycomb lattices. 

\begin{figure*}[htpb]
	\centerline{
		\includegraphics[width=\textwidth]{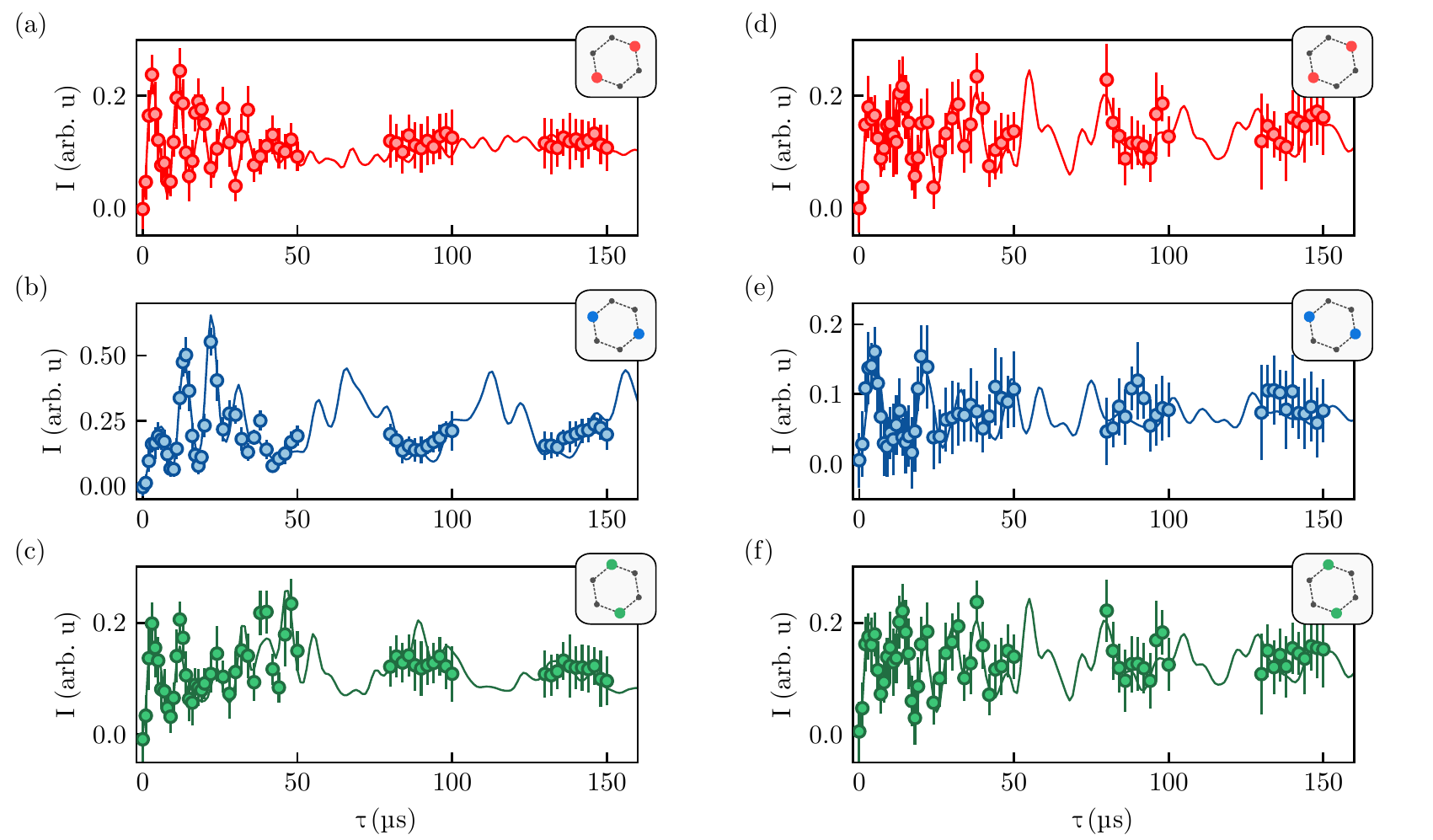}
	}
	\caption{Bragg scattering of a molecular BEC exposed to an optical lattice with triangular (a, b, c) and honeycomb (d, e, f) geometries. Data of Figs. \ref{Fig:BraggT} and \ref{Fig:BraggH} for the full experimentally probed time domain. Population ($I$) as a function of pulse duration $\tau$ in the six first order diffraction peaks. Experimental data points averaged over 15 realizations are given with error bars representing one standard deviation. Corresponding peaks are indicated by the top-right insets. The solid lines show the simulated population obtained from a fit to the model.}
	\label{Fig:BraggHT_Full}
\end{figure*}

Building on this analysis, we adjusted the experimental parameters of the Bragg experiments as discussed in Sec.~\ref{sec:lattice} and obtained the results shown in Figs.\,\ref{Fig:BraggT} and \ref{Fig:BraggH}, where we find excellent quantitative agreement between the experiments and our prediction over the whole diffraction pattern dynamics. Fig.\, \ref{Fig:BraggHT_Full} shows the data of Figs. \ref{Fig:BraggT} and \ref{Fig:BraggH} plotted over an extended time domain. Fits based on the model are obtained by allowing the two-photon Rabi frequencies $\Omega_{12}$, $\Omega_{13}$, $\Omega_{23}$ and a relaxation coefficient as fitting parameters, from which we extract the intensity of each lattice beam. The fit for the two-photon Rabi frequencies yields $\Omega_{12} = 2\pi \times 44.8\:(38.9)$\,kHz, $\Omega_{13}= 2\pi \times 65.0\:(64.4)\,$kHz, and $\Omega_{23} =2\pi \times 59.8\:(65.3)\,$kHz for the triangular (resp. honeycomb) lattice. For the first 50\,ms, the model gives coefficients of determination of $R^2 = 0.81$ and $0.69$ for the triangular and honeycomb configurations, respectively. Over the full range of the dynamics we obtain $R^2 = 0.73$ and $0.57$, respectively.

The relaxation coefficient accounts for various sources of decoherence that can arise during the Bragg dynamics. An obvious source of decoherence stems from the inhomogeneity of the lattice beams when the spatial extent of the molecular BEC is not negligible compared to the beam sizes \cite{Denschlag2002}. In our fitting procedure, we therefore encode the relaxation by using an effective size of the molecular BEC as a fitting parameter. The fitted cloud sizes are $10.6(4)\,\mu$m and $13(1)\,\mu$m for the triangular and honeycomb lattices, respectively. These can be translated into a characteristic relaxation time of $\eta \simeq 35\,\mu$s (resp. $\eta \simeq 50\,\mu$s) for the triangular (resp. honeycomb) lattice, where $\eta$ enters as a damping time of a Gaussian envelope $\exp[-t^2/(2\eta^2)]$. We find that the fitted cloud size is 50\% (resp. 80\%) larger than the measured size for the triangular (resp. honeycomb) lattice. This indicates that the lattice inhomogeneity is indeed an important source of decoherence, but also signals that other decohering processes are at play. Other sources of decoherence include residual dimer-dimer collisions. The most likely additional contribution is, however, a measurement artifact resulting from shot-to-shot lattice beam intensity fluctuations, which leads to an apparent decoherence. For instance, an intensity fluctuation of 5\% would yield a 100\,$\mu$s relaxation time.

While the analysis above shows that we have an accurate description of the Bragg dynamics, the strong differences between the diffraction patterns resulting from the triangular and honeycomb lattices may nevertheless seem counterintuitive. Indeed, under inspection of the full Hamiltonian above, one can see that the two cases only differ by a minus sign in the expression of the matrix elements $\left\langle \alpha \right| V \left| \beta \right\rangle$. To build intuition on the importance of this sign difference and the role it can play in the dynamics, we have developed a simplified model that we describe below.

The full Hamiltonian above can be greatly simplified under two assumptions. Firstly, we assume that all three lattice arms have equal intensity, such that we can write $\Omega_1 = ... = \Omega_6 = \Omega$ (chosen to be a real number for simplicity). For the triangular lattice, all polarizations are parallel to each other, leading to a negative Rabi frequency $\Omega_{\mathrm{triangular}} < 0$, while for the honeycomb lattice the relative angle between polarization vectors lead to a sign reversal $\Omega_{\mathrm{honeycomb}} = -\Omega_{\mathrm{triangular}}/2 > 0$. Secondly, if we take the Rabi frequency $\Omega$ to be small compared to the kinetic energy associated with the first non-zero momentum states $\Delta$, the occupation of large momentum states is suppressed due to the off-resonant nature of the Raman process. To the lowest approximation order we can therefore only consider the first $6$ non-zero momentum states $(\mathbf{b}_1, ... , \mathbf{b}_6)$, which have the same kinetic energy $\hbar \Delta > 0$ and therefore couple  resonantly to each other. In the $(\mathbf{b}_0,\mathbf{b}_1, ..., \mathbf{b}_6)$ basis, the Hamiltonian is then:
\begingroup
\renewcommand*{\arraystretch}{1.25}
\begin{equation*}
H = \hbar \begin{pmatrix}
0 &\frac{\Omega}{2} &\frac{\Omega}{2} &\frac{\Omega}{2} &\frac{\Omega}{2} &\frac{\Omega}{2} &\frac{\Omega}{2}\\
\frac{\Omega}{2} &\Delta &\frac{\Omega}{2} &0 &0 &0  &\frac{\Omega}{2}\\
\frac{\Omega}{2} &\frac{\Omega}{2} &\Delta &\frac{\Omega}{2} &0  &0 &0\\
\frac{\Omega}{2} &0 &\frac{\Omega}{2} &\Delta &\frac{\Omega}{2} &0  &0\\
\frac{\Omega}{2} &0 &0 &\frac{\Omega}{2} &\Delta &\frac{\Omega}{2} &0\\
\frac{\Omega}{2} &0 &0 &0 &\frac{\Omega}{2} &\Delta &\frac{\Omega}{2} \\
\frac{\Omega}{2} &\frac{\Omega}{2} &0 &0 &0 &\frac{\Omega}{2} &\Delta
\end{pmatrix}.
\end{equation*} 
\endgroup
Due to the coupling, however, the initial basis $(\mathbf{b}_1,...,\mathbf{b}_6)$ of momentum states is not the most adapted to diagonalize the Hamiltonian. We therefore simplify $H$ further by performing the basis change $(\mathbf{b}_0,\mathbf{b}_1, ..., \mathbf{b}_6) \to (\mathbf{b}_0,\mathbf{\tilde{b}}_1, ..., \mathbf{\tilde{b}}_6)$, with $\mathbf{\tilde{b}}_k = \sqrt{1/6} \sum_{j=1}^6 e^{i\frac{2\pi(k-1)j}{6}} \mathbf{b}_j$. This makes the Hamiltonian diagonal in the subspace $(\mathbf{\tilde{b}}_1,...,\mathbf{\tilde{b}}_6)$ with associated energies $E_k/\hbar = \Delta + \Omega\, \mathrm{cos}\left(\frac{2\pi(k-1)}{6}\right)$. Furthermore, the new coupling constants $\tilde{\Omega}_k = \sqrt{1/6} \sum_{j=1}^6 e^{i\frac{2\pi(k-1)j}{6}} \Omega$ are all zero except $\tilde{\Omega}_1 = \sqrt{6} \Omega$. The Hamiltonian $\tilde{H}$ in this new basis becomes:
\begin{small}
\begin{equation*}
\begin{split}
&\tilde{H}/\hbar  =\\
& \begin{pmatrix}
0 &\sqrt{6} \frac{\Omega}{2} &0 &0 &0 &0 &0\\
\sqrt{6} \frac{\Omega}{2} &\Delta + \Omega &0 &0 &0 &0  &0\\
0 &0 &\Delta + \frac{\Omega}{2} &0 &0  &0 &0\\
0 &0 &0 &\Delta - \frac{\Omega}{2} &0 &0  &0\\
0 &0 &0 &0 &\Delta - \Omega &0 &0\\
0 &0 &0 &0 &0 &\Delta - \frac{\Omega}{2} &0 \\
0 &0 &0 &0 &0 &0 &\Delta + \frac{\Omega}{2}
\end{pmatrix}
\end{split}.
\end{equation*}
\end{small}

The Hamiltonian is thus reduced to a two-level system with an effective detuning $\tilde{\Delta} = \Delta+\Omega$ and an effective Rabi frequency $\tilde{\Omega} = \sqrt{6} \Omega$, meaning that, when starting with all atoms in $\left|\mathbf{p=0}\right\rangle$, the system will oscillate between the initial state and the symmetric superposition of the $6$ first excited momentum states. The difference between the triangular and honeycomb lattices (for a given coupling strength $\left|\Omega\right|$) hence comes from the sign of $\Omega$. For the triangular lattice ($\Omega < 0$), the resonant coupling of $(\mathbf{b}_1,...,\mathbf{b}_6)$ brings the excited state closer to resonance, whereas for the honeycomb lattice ($\Omega > 0$) it pushes the excited states further from resonance. For a given lattice depth, we thus expect the triangular lattice to transfer a larger fraction of atoms to excited states than the honeycomb lattice. This qualitative difference is a generic feature and holds beyond the low intensity limit and the balanced intensity case, as is observed in the diffraction patterns in Figs. \ref{Fig:BraggT} and \ref{Fig:BraggH}.

As a validation of this simplified model, we perform Bragg scattering measurements in a triangular lattice at low laser power in order to approach the condition $\Omega\ll\Delta$. The results are shown in Fig.\,\ref{Fig:Bragg}, where the diffraction patterns display a predominant population of only the central peak and the first non-zero momentum states, as expected in the low-intensity regime. From these diffraction patterns, we extract two populations: the one of the zeroth order and the summed population of all first order diffraction peaks, which we plot as a function of $\tau$. We observe that the time-evolution of the populations is well fitted with damped Rabi oscillations, confirming the two-level picture. From the fit, we obtain an effective Rabi frequency $|\tilde{\Omega}| \simeq 2\pi\times 64.3(5)\,\mathrm{kHz}$ and an effective detuning $\tilde{\Delta} \simeq 2\pi \times 10(3)\,\mathrm{kHz}$, which are in agreement with the predicted values $|\tilde{\Omega}| = \sqrt{6}|\Omega| \simeq 2\pi \times 68(5)\,\mathrm{kHz}$ and $\tilde{\Delta} = \Delta + \Omega \simeq 15(2)\,\mathrm{kHz}$, where $|\Omega| \simeq 2 \pi \times 28(2)\,\mathrm{kHz}$ is obtained from an independent calibration and $\Delta=E_{\rm L}/\hbar=2\pi\times43$\,kHz directly stems from the lattice geometry. We attribute the slight discrepancy on $\tilde{\Delta}$ to the fact that the two conditions required for the simplified model are not perfectly fulfilled. We also extract a relaxation time of $59(6)\,\mu$s form the fit.

\providecommand{\noopsort}[1]{}\providecommand{\singleletter}[1]{#1}%


\begin{thebibliography}{109}
\bibitem [{ToB()}]{ToBePub2023}%
  \BibitemOpen
  \href@noop {} {\bibinfo  {journal} {To be published}\ }\BibitemShut {NoStop}%
\end{thebibliography}
\end{document}